\documentclass[%
 aip,
 amsmath,amssymb,
reprint,%
]{revtex4-1}

\usepackage{graphicx}
\usepackage{subcaption} 
\usepackage{dcolumn}
\usepackage{bm}
\usepackage[usenames,dvipsnames]{xcolor}
\usepackage[utf8]{inputenc}
\usepackage{textcomp}
\usepackage[T1]{fontenc}
\usepackage{mathptmx}
\usepackage{textcomp}
\usepackage{etoolbox}
\usepackage{natbib}
\usepackage[hidelinks]{hyperref}
\usepackage{array}
\usepackage{multirow}
\usepackage{ulem}
\usepackage{orcidlink}

\usepackage[
lined,
ruled, 
vlined,
commentsnumbered, 
]{algorithm2e}
\usepackage{algpseudocode}

\makeatletter
\def\@email#1#2{%
 \endgroup
 \patchcmd{\titleblock@produce}
  {\frontmatter@RRAPformat}
  {\frontmatter@RRAPformat{\produce@RRAP{*#1\href{mailto:#2}{#2}}}\frontmatter@RRAPformat}
  {}{}
}%
\makeatother

\newcommand{\eqsplit}[1]{\begin{equation} \begin{split} #1 \end{split}\end{equation}}

\newcommand{\norb}{L}
\newcommand{\nimp}{L_{\text{imp}}}

\newcommand{\Tr}{\text{Tr}}

\newcommand{\hartree}{\ E_{\text{h}}}
\newcommand{\bohr}{\ a_0}

\newcommand{\dchemistry}{Department of Chemistry and Chemical Biology, Rutgers University, Piscataway, NJ 08854, USA}
\newcommand{\columbia}{Department of Chemistry, Columbia University, New York, NY 10027, USA}

\begin{document}
\preprint{AIP/123-QED}

\title{Ab initio quantum embedding at finite temperature with density matrix embedding theory}

\author{Laurence Giordano\orcidlink{https://orcid.org/0000-0002-4624-4444}}
\affiliation{\dchemistry}
\author{Y. Stanley Tan\orcidlink{https://orcid.org/0009-0006-7926-7278}}
\affiliation{\dchemistry}
\author{Zhi-Hao Cui\orcidlink{https://https://orcid.org/0000-0002-7389-4063}}
\affiliation{\columbia}
\author{Chong Sun\orcidlink{https://orcid.org/0000-0002-8299-9094}}
\affiliation{\dchemistry}
\email{chongs0419@gmail.com}

\date{\today}

\begin{abstract}
We present a finite-temperature extension of density matrix embedding theory (FT-DMET) for realistic crystalline systems. We describe a practical framework for constructing extended bath orbitals, solving the embedding problem, and performing DMET self-consistency at finite temperature. To reduce computational cost, we introduce strategies based on mutual-information-guided bath truncation, controlled treatment of the thermal electron number without explicit optimization, and the use of low-temperature impurity solvers and one-shot FT-DMET in the low-temperature regime. We apply this approach to periodic hydrogen chains and square lattices to characterize their finite-temperature phases. We observe the Pomeranchuk-like effect in one dimension and enhanced stability of long-range order in two dimensions.
\end{abstract}

\maketitle

\section{\label{sec:hlatt_intro}Introduction}
Finite-temperature simulation of realistic materials remains a grand challenge due to the inherent complexity of thermal states compared to the ground state. While a ground state typically resides on a low-dimensional manifold of the Hilbert space, a finite-temperature state spans a much larger domain of the Fock space, requiring substantially more information to capture the interplay between electronic correlations and entropic effects~\cite{bennett1996mixed, vedral2002entropy, plbnio2007entanglement, amico2008entanglement, sun2021thesis}.
Accurate modeling of these states is crucial for understanding phenomena that emerge primarily under thermal excitation, including fundamental phase transitions such as the N\'{e}el transition in magnetic insulators and the thermal Mott transition in transition-metal oxides~\cite{blundell2001magnetism, sachdev2011quantum}. Thermal simulations also probe exotic regimes where quantum and thermal fluctuations coexist, including the pseudogap and strange metal phases of high-temperature superconductors~\cite{orenstein2000advances, zhou2021high, timusk1999pseudogap, greene2020strange} and the breakdown of thermalization in many-body localized systems~\cite{ogsanesyan2007localization, huse2013localization, sun2024electron}. 
Consequently, developing numerical methods capable of efficiently compressing this vast state space is essential for predicting the thermodynamic properties of strongly correlated materials.

Quantum embedding methods strike a balance between accuracy and scalability, making them effective for studying extended, strongly correlated systems~\cite{sun2016quantum, georges1996dynamical, 
lechermann2007risb,
knizia2012dmet_prl, lan2016rigorous, zhu2020efficient, cui2020efficient}. Among these, density matrix embedding theory (DMET) is a wavefunction-in-wavefunction embedding scheme that maps the full system onto a localized impurity coupled to a discrete bath~\cite{knizia2012dmet_prl, knizia2013dmet_jctc, wouters2016practical, wouters2017fiveyears}. DMET employs an effective one-body correlation potential, optimized self-consistently to align the local one-particle reduced density matrix (1RDM) of a low-level mean-field state with that of a high-level embedding solution. This simple yet powerful formalism makes DMET well suited for simulating strongly correlated phenomena in both model systems and realistic materials~\cite{zheng2016ground, zheng2017stripe, cui2020threeband, cui2022cuprate, cui2025abinitio}.

The success of DMET stems from its ability to represent the full system through a compact embedding problem—comprising the impurity and a bath—that captures the entanglement between the impurity and its environment at a ``correlated'' mean-field level. By leveraging the Schmidt decomposition, DMET rigorously compresses the environment into a bath of the same size as the impurity, achieving high accuracy at a low computational cost.
At finite temperature, however, thermal fluctuations broaden the correlations between the impurity and its environment. The resulting transition from an entanglement area law to a volume law necessitates a bath space larger than the standard zero-temperature construction.
In previous work~\cite{sun2020ftdmet}, we introduced finite-temperature DMET (FT-DMET) with an extended bath constructed from a moment expansion of the mean-field density matrix. This approach preserves most of the impurity-environment entanglement while keeping the bath compact and was validated on one- and two-dimensional Hubbard models.

In this work, we extend the FT-DMET method to \textit{ab initio} systems and provide a practical framework for its implementation. A key bottleneck in DMET is the accurate solution of the embedding problem, whose cost increases at finite temperature. To address this, we develop strategies to compress the embedding space and reduce the number of impurity solver calls.
Specifically, we introduce a protocol for constructing bath orbitals using mutual information analysis, enabling systematic truncation of the expanded thermal environment. We also present methods for approximating the chemical potential in grand-canonical simulations and a low-temperature formula to reduce computational overhead. We apply this framework to periodic hydrogen chains and square lattices, revealing a rich landscape of finite-temperature phases. By analyzing magnetic moments, spin-spin correlations, double occupancy, and dimerization, we characterize the interplay between magnetic order and thermal entropy. In particular, we observe the Pomeranchuk effect~\cite{richardson1997pomeranchuck} in the one-dimensional chain and enhanced stability of long-range order in the two-dimensional lattice.

The remainder of the paper is organized as follows. In Section~\ref{sec:method}, we discuss the framework in detail, including modeling the full system with periodic boundary conditions (PBC), constructing the finite-temperature bath, solving the impurity problem within a grand-canonical formalism, and implementing DMET self-consistency with the low-temperature approximation. Section~\ref{sec:result} presents applications to hydrogen chains and square lattices, and Section~\ref{sec:conclusion} concludes with a summary and outlook.

\section{Theory\label{sec:method}}

\subsection{Modeling a crystalline system\label{sec:method_lattice}}
We begin by modeling a crystalline system under periodic boundary conditions (PBC) by defining a unit cell in real space and a $k$-point mesh in reciprocal space. One unit cell is selected as the impurity (fragment), while the remaining lattice is treated as the environment. For three-dimensional systems, the unit cell may be either the primitive cell or an enlarged supercell. For lower-dimensional systems, we still adopt a three-dimensional unit cell because the electron distribution along the nonperiodic direction is nonzero, while $k$-points are sampled only along the periodic directions. The \textsc{PySCF} package~\cite{sun2017pyscf,sun2020pyscf} is used to initialize the lattice and perform mean-field calculations to obtain the Fock matrix and molecular orbitals (MO).

Defining the impurity requires further localization of the atomic orbitals (AO). We use intrinsic and projected atomic orbitals (IAO+PAO)~\citep{knizia2013intrinsic, Saebo1993local} as the localized orbitals (LO). An adapted IAO procedure maps the occupied MOs onto a set of LOs, while the PAOs, orthogonalized by L\"{o}wdin’s method~\cite{aiken1980on}, span the virtual space. 
This IAO+PAO strategy is now standard in \textit{ab initio} quantum embedding simulations~\cite{cui2020efficient, zhu2020efficient, cui2025abinitio, zhu2025towards}. Since the remainder of the DMET calculations are performed in the IAO+PAO basis, we refer to this basis as the AO basis in the following discussion for simplicity.

Because the full lattice is treated at the mean-field level, an auxiliary one-body potential acting on the impurity is introduced to calibrate the impurity-environment interaction.
This potential is referred to as the \textit{correlation potential} and is denoted by $V_c$ in this paper. Owing to the translational invariance of the crystalline system, $V_c$ is added to each unit cell of the lattice. The mean-field solution is therefore obtained by solving the self-consistent field (SCF) problem with a modified Fock operator
\eqsplit{\label{eq:fock_w_corr}
\tilde{F} = F + \sum_{x} V_c^x,
}
where $F$ is the Fock operator of the original lattice Hamiltonian, and $x$ runs over all unit cells. The correlation potential is determined through the DMET self-consistency procedure, which is briefly described in Section~\ref{sec:ftdmet_loop}.

The solution to Eq.~\eqref{eq:fock_w_corr} is obtained using finite-temperature Hartree-Fock (FT-HF). FT-HF follows the same procedure as ground-state HF, with two modifications. First, the one-particle reduced density matrix (1RDM), $D_{\text{HF}}(\beta)$,  is constructed with the Fermi-Dirac distribution, 
\eqsplit{\label{eq:fermi_dirac}
\text{Occupancy: }&\quad f(\varepsilon_i) = \left[1+e^{\beta(\varepsilon_i-\mu)}\right]^{-1}, \\
\text{1RDM: }&\quad D_{\text{HF}}(\beta) = \left[ 1 + e^{\beta(\tilde{F} - \mu I)}\right]^{-1},
}
where $\beta = 1/k_\text{B}T$ is the inverse temperature, $\varepsilon_i$ is the energy of the $i$-th MO, $\mu$ is the chemical potential, and $I$ is an identity matrix of the same dimension as $\tilde{F}$. Throughout this work, the Boltzmann constant is set to $k_\text{B}=1$, such that $\beta = 1/T$.
In the FT-HF self-consistency loop, $D_{\text{HF}}(\beta)$ updates $\tilde{F}$, rendering the Fock matrix temperature dependent. 
Since only the original Fock matrix $F$ requires self-consistent optimization, $F$ is first solved once, after which $\tilde{F}$ is constructed using the $V_c$ obtained at each DMET iteration, and Eq.~\eqref{eq:fermi_dirac} is solved without further updating $\tilde{F}$.

The second modification involves adjusting the chemical potential $\mu$ to ensure the correct total electron number,
\eqsplit{
\sum_{\alpha=\uparrow\downarrow}\sum_{i=1}^{L} f(\varepsilon^\alpha_i; \mu) = N_e
}
where $\alpha$ represents the spin degree of freedom. The same $\mu$ value is used for both spin channels. The converged $D_{\text{HF}}(\beta)$ is then used to construct the bath orbitals, as discussed in the next section.

\begin{figure*}[t!]
\begin{subcaptionbox}{Full-impurity bath\label{sub:bath_full}}{\includegraphics[width=0.207\linewidth]{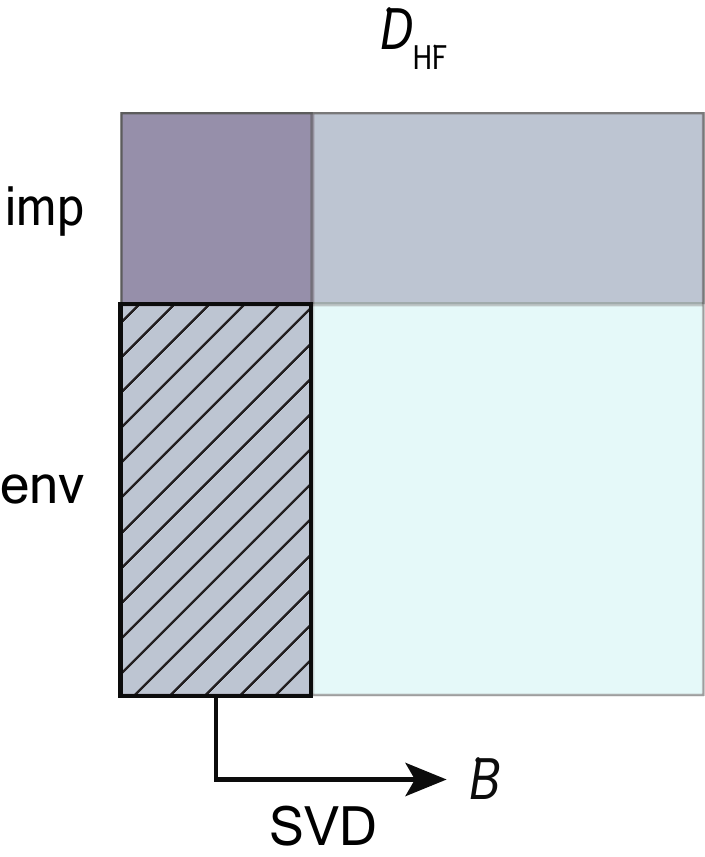}}
\end{subcaptionbox}
\hfill
\begin{subcaptionbox}{Moment-expansion bath\label{sub:bath_moment}}{\includegraphics[width=0.402\linewidth]{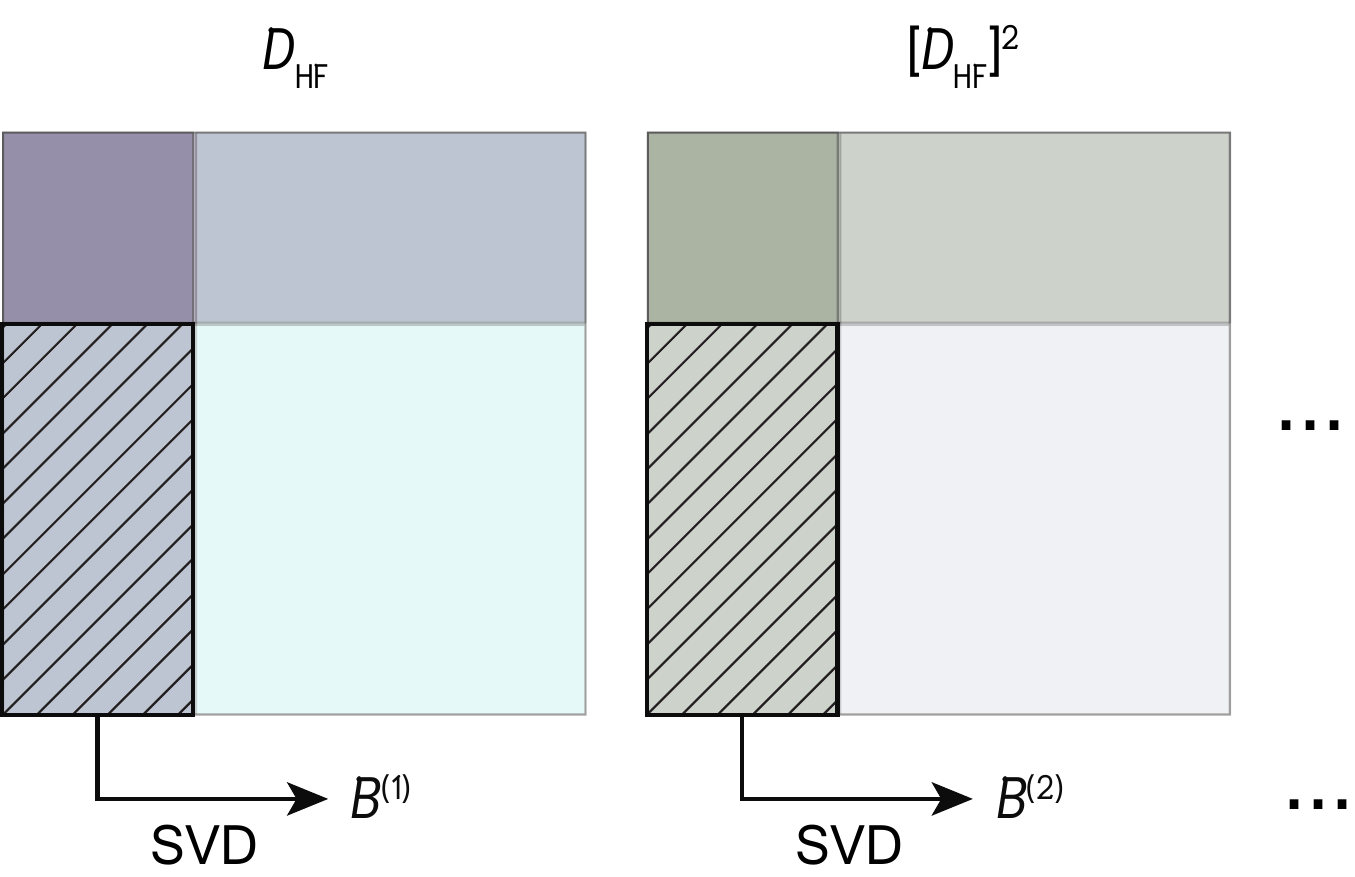}}
\end{subcaptionbox}
\hfill
\begin{subcaptionbox}{Valence-only bath\label{sub:bath_valence}}{\includegraphics[width=0.172\linewidth]{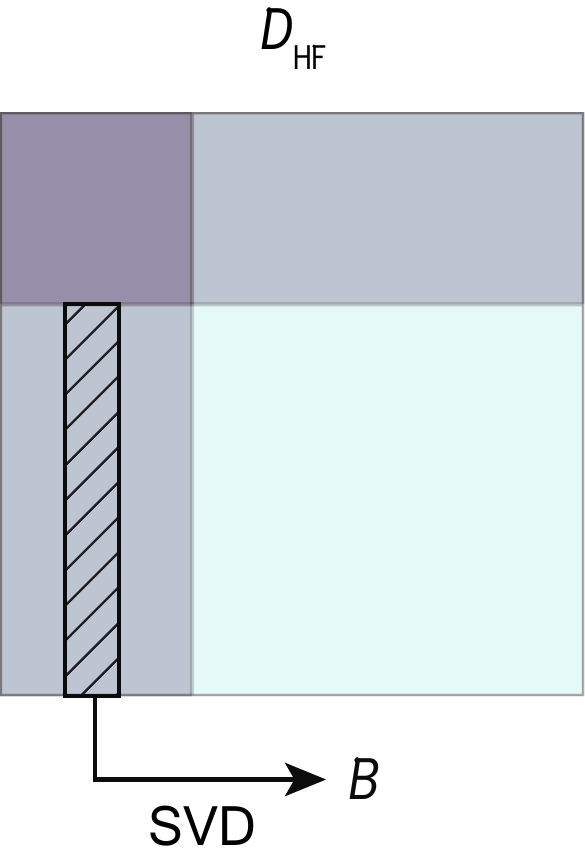}}
\end{subcaptionbox}

\caption{Strategies for constructing the DMET bath. The shaded impurity-environment blocks (or the valence-environment block in panel (c)) are used to perform the SVD from which the bath orbitals are derived.}
\label{fig:bath_strategies}
\end{figure*}

\subsection{Obtaining bath orbitals~\label{sec:method_bath}}
The bath represents the portion of the environment that is entangled with the impurity. For the ground state, the bath dimension is at most the size of the impurity space, reflecting the area law of entanglement entropy~\cite{eisert2010arealaw}. The DMET bath captures the mean-field entanglement between the impurity and the environment and can be obtained from a singular value decomposition (SVD) of the impurity-environment block of the Hartree-Fock 1RDM,
\eqsplit{\label{eq:gs_bath_svd}
D_{\text{HF}}^{\text{env-imp}} = B\lambda U,
}
where the columns of $D_{\text{HF}}^{\text{env-imp}}$ corresponds to impurity orbitals and the rows correspond to environment orbitals. The columns of $B$ define the bath orbitals expressed in the AO basis, as illustrated in FIG. \ref{sub:bath_full}.

At finite temperature, however, the entanglement entropy follows a volume law, and the bath space is generally larger than in the ground-state case. In our previous work~\cite{sun2020ftdmet}, we derived a \textit{moment expansion} procedure to construct finite-temperature bath orbitals, as shown in FIG.~\ref{sub:bath_moment}. 
By applying Eq.~\eqref{eq:gs_bath_svd} to successive moments of $D_{\text{HF}}(\beta)$, a sequence of bath orbitals is generated, and the finite-temperature bath space is defined as the span of these orbitals:
\eqsplit{\label{eq:ft_bath}
&\left[D_{\text{HF}}(\beta)\right]^m \rightarrow B^{(m)}, \quad m = 1, 2, \cdots, \\
&B = [B^{(1)}, B^{(2)}, \cdots].
}
The resulting bath orbitals are orthonormalized, and linear dependencies are removed. In practice, truncating the expansion at second or third order provides an accurate approximation to the exact finite-temperature bath.

The bath constructions described above are effective when all impurity orbitals are entangled with the environment. However, in most \textit{ab initio} systems, the impurity contains core, valence, and virtual orbitals, and the low-lying core orbitals as well as the high-energy virtual orbitals exhibit negligible entanglement with the environment. Enforcing the bath space to match the full impurity dimension therefore introduces redundancy into the embedding problem, which unnecessarily enlarges the embedding space and complicates convergence in the DMET self-consistency procedure.

To remove this redundancy, we divide the impurity orbitals into core, valence, and virtual subspaces, and generate bath orbitals only for the valence subspace~\cite{sun2014exact, wouters2016practical, cui2020efficient}. In this approach, bath orbitals are obtained by performing SVD on the valence-environment block of the mean-field 1RDM, and the number of bath orbitals is at most the number of valence impurity orbitals $L_{\text{val}}$, as illustrated in FIG. \ref{sub:bath_valence}.

\begin{figure}[t!]
    \centering
    \begin{subcaptionbox}{Mutual information vs. bath size\label{sub:mi_bathzise}}{\includegraphics[width=\linewidth]{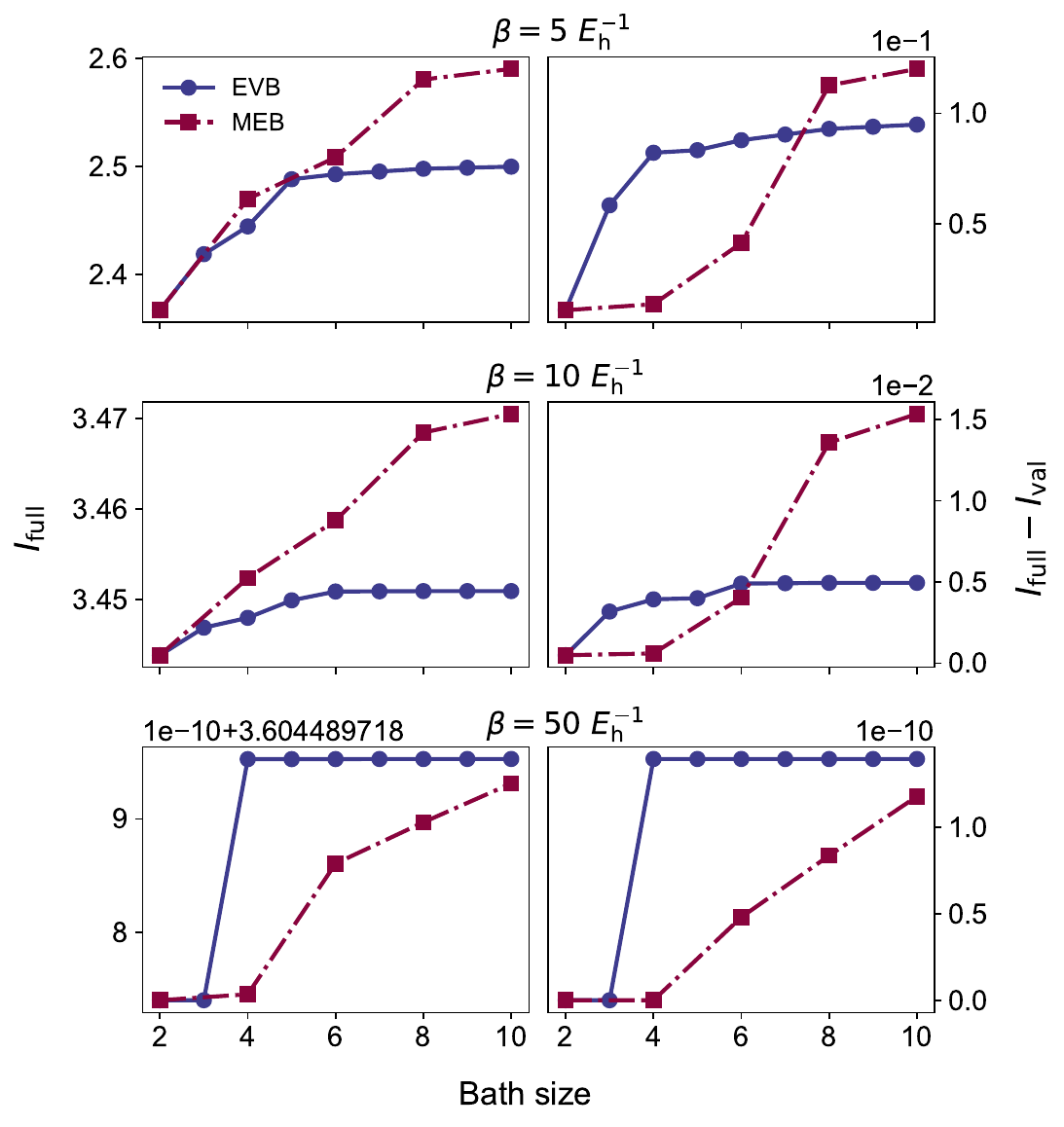}}
\end{subcaptionbox}

\vspace{0.5em}
\begin{subcaptionbox}{Mutual information vs. $\beta$ \label{sub:mi_beta}}{\includegraphics[width=\linewidth]{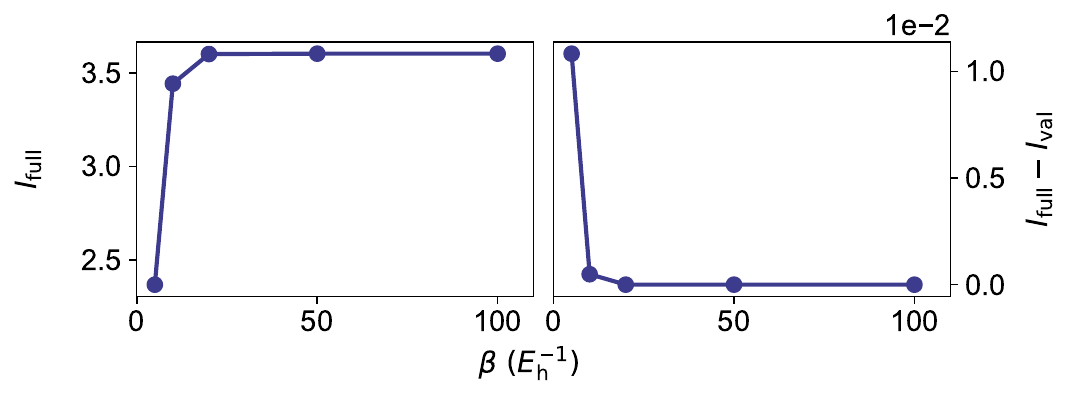}}
\end{subcaptionbox}
    \caption{Mutual information between the impurity and bath for a hydrogen chain with 10 impurity orbitals, including 2 valence orbitals. $I_\text{full}$ denotes the mutual information between the full impurity and bath, and $I_\text{val}$ between the valence impurity and bath. Shown are  $I_\text{full}$ and $I_\text{full} - I_\text{val}$  versus (a) bath size at several $\beta$ values and (b) inverse temperature $\beta$ with the bath size fixed at 2 orbitals.
    }
    \label{fig:mutual_info}
\end{figure}

\subsubsection{Ab initio bath at finite temperature} 
At finite temperature, portions of the core and virtual impurity spaces begin to entangle with the environment. In addition, the transition of the entanglement entropy from an area law to a volume law leads to an increased required bath space~\cite{sun2020ftdmet}. Based on these observations, we propose two strategies for constructing a finite-temperature \textit{ab initio} bath.

The first strategy constructs the bath using all impurity orbitals and subsequently truncates the bath space based on the singular values obtained from the SVD.
The second strategy constructs the bath using only the valence impurity orbitals and applies the moment-expansion procedure described in Eq.~\eqref{eq:ft_bath}.
We refer to the first strategy as the extended-valence bath (EVB) approach and the second as the moment-expansion bath (MEB) approach. We compare these two approaches by evaluating the entanglement between the bath and (i) the full impurity and (ii) the valence impurity.

The entanglement is quantified using the mutual information between two subsystems $A$ and $B$, defined as
\eqsplit{\label{eq:mutual_info}
I({A}; {B}) = S_{A} + S_{B} - S_{AB},
}
where $S_{X} = -\Tr [\rho_X \log \rho_X]$ is the von Neumann entropy of system $X$, and $\rho_X$ is the corresponding density matrix. 

Since the bath orbitals are defined at the mean-field level, we adopt a one-particle proxy of Eq.~\eqref{eq:mutual_info} and replace the density matrix $\rho_X$ with the mean-field 1RDM:
\eqsplit{
D_X = P_X D_{\mathrm{HF}} P_X ,
}
where $P_X$ is the projection operator onto subsystem $X$.

Correspondingly, we employ a one-particle reduced entropy,
\eqsplit{
S_X^{(1)} = -\Tr\left[ D_X \log D_X + (I_X - D_X)\log (I_X - D_X) \right],
}
where $I_X$ is the identity matrix with the same dimension as $D_X$.

We denote the mutual information between the full impurity and the bath as $I_{\mathrm{full}}$, and that between the valence impurity and the bath as $I_{\mathrm{val}}$.
FIG.~\ref{fig:mutual_info} presents the analysis for a periodic hydrogen chain at an interatomic distance $R = 1.5$ bohr ($1.5\bohr$), where the impurity consists of two hydrogen atoms. A $[1,1,3]$ $k$-point mesh is used. With the cc-pVDZ basis, the impurity contains 10 orbitals, of which 2 are valence orbitals.


In FIG.~\ref{sub:mi_bathzise}, we examine $I_{\mathrm{full}}$ and $I_{\mathrm{full}} - I_{\mathrm{val}}$ as functions of the bath size for both the EVB and MEB approaches.
At very low temperature, the entanglement resembles ground-state behavior, and only the valence impurity orbitals contribute appreciably to the impurity-environment entanglement. As the temperature increases, the embedding space becomes a mixed state and the core impurity orbitals begin to entangle with the environment. The crossing of the EVB and MEB curves indicates that the leading correction to the valence-only bath first arises from the core impurity orbitals, after which thermal fluctuations require the inclusion of higher-order bath orbitals generated by the 1RDM moment expansion.

{The slow convergence of the MEB curves in FIG.~\ref{sub:mi_bathzise} at elevated temperatures reflects the expected behavior of impurity-environment entanglement: with increasing temperature, entanglement becomes increasingly delocalized and approaches a maximum in the infinite-temperature limit. Consequently, a larger number of bath orbitals are, in principle, entangled with the impurity, a feature not captured by the EVB construction, which exhibits apparent early convergence with a small bath size. This delocalized entanglement, however, does not imply a breakdown of DMET at high temperatures. As temperature increases, quantum correlations become progressively less important and the mean-field description becomes increasingly accurate. In this regime, the detailed impurity-environment coupling plays a reduced role in determining thermal observables, and the dependence on bath size correspondingly diminishes. This interplay between thermal fluctuations and quantum effects underlies the robustness of FT-DMET across the full temperature range~\cite{sun2020ftdmet}.}

In FIG.~\ref{sub:mi_beta}, we plot the same quantities as functions of the inverse temperature $\beta$ using the valence-only bath. This benchmark identifies the temperature regime in which the valence-only bath remains a reasonable approximation of the impurity-environment entanglement. For the system studied here, $\beta > 10 \hartree^{-1}$ may be regarded as the low-temperature regime.

The practical conclusions are summarized below:
\begin{enumerate}
\item At low temperature, the valence-only bath captures most of the impurity-environment entanglement.
\item At higher temperature, the MEB approach systematically improves the bath quality, whereas the EVB approach saturates after adding only a small number of additional bath orbitals.
\end{enumerate}

\begin{figure}[t!]
    \centering
    \includegraphics[width=0.55\linewidth]{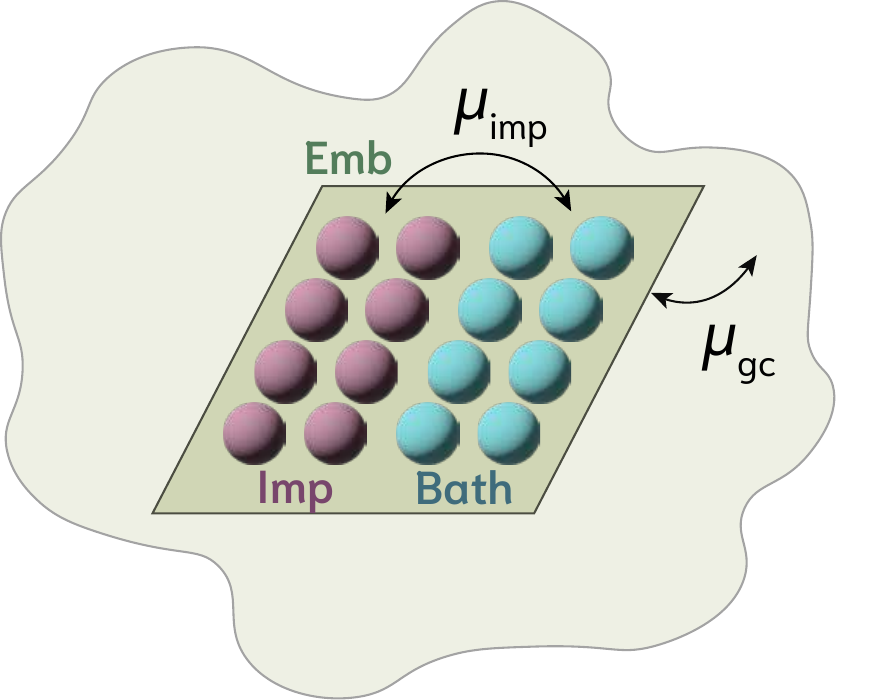}
    \caption{Enforcing the correct electron distribution in the embedding and impurity spaces using two chemical potentials, $\mu_\text{gc}$ and $\mu_{\text{imp}}$.}
    \label{fig:two_mu_values}
\end{figure}


\begin{figure*}[t!]
    \centering
    \includegraphics[width=0.8\linewidth]{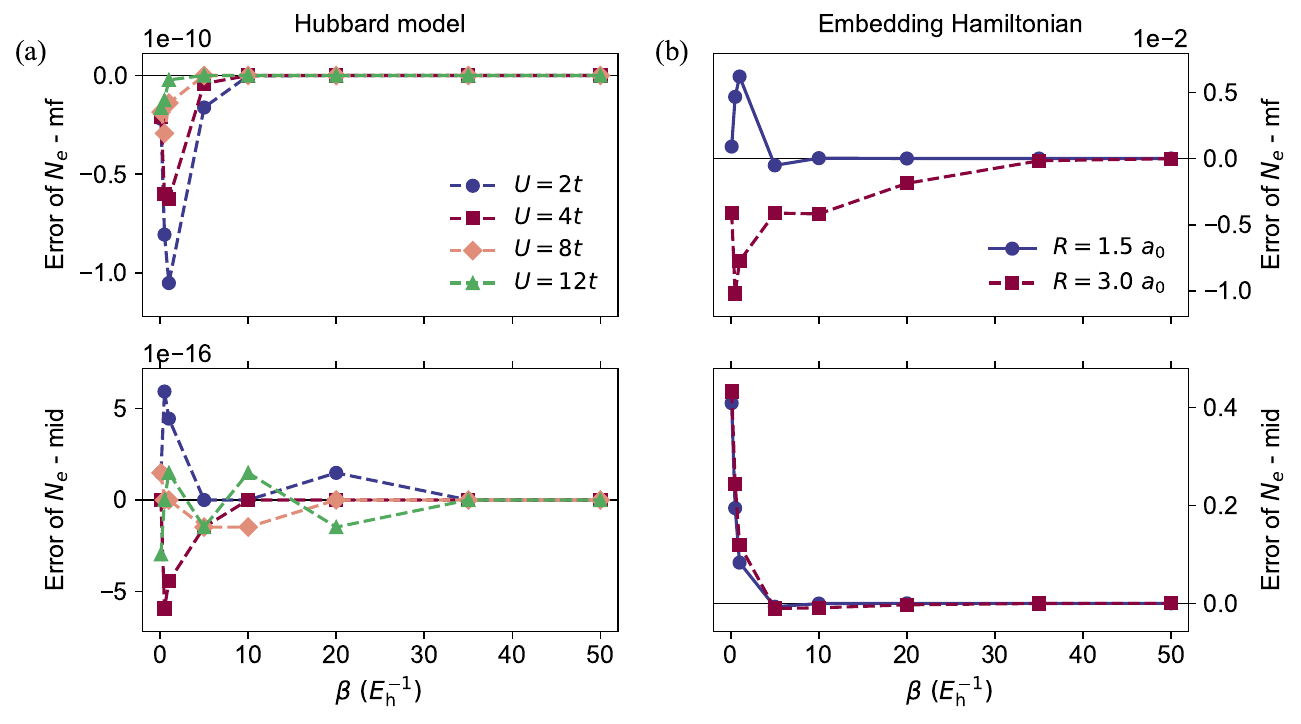}
\caption{Error in the thermal-average electron number evaluated using the mean-field chemical potential $\mu_\text{gc}^\text{mf}$ (upper panels) and the mid-gap chemical potential $\mu_\text{gc}^\text{mid}$ (lower panels) for (a) a one-dimensional (6e, 6o) Hubbard model and (b) a (4e, 6o) embedding Hamiltonian derived from a hydrogen chain.}
    \label{fig:mu_guess_nelectron}
\end{figure*}

\subsection{Solving the embedding problem\label{sec:embedding_problem}}
The embedding Hamiltonian is defined as the projection of the full lattice Hamiltonian onto the embedding space (impurity plus bath). Two standard approaches exist:
(1) the non-interacting bath (NIB) formalism and (2) the interacting bath (IB) formalism~\cite{wouters2016practical}.
In the NIB formalism, only two-body interactions within the impurity are treated explicitly, while impurity-bath and bath-bath interactions are approximated using the projected mean-field Fock terms ($J$ and $K$) and the correlation potential. 
In contrast, the IB formalism projects the full lattice Hamiltonian onto the embedding space, which retains a more accurate description of all two-body interactions at the cost of increased computational effort.

Because Coulomb interactions in \textit{ab initio} systems are typically delocalized, we adopt the IB formalism to preserve as much of the two-body physics as possible. The IB embedding Hamiltonian is given by
\begin{equation}
{H}_{\text{emb}} = \sum_{pq\in \text{emb}}h^{\text{emb}}_{pq}c^{\dag}_pc_q + \frac{1}{2}\sum_{pqrs\in \text{emb}}\left(pq|rs\right)c^{\dag}_pc^{\dag}_r
c_sc_q.
\end{equation}

Projecting the four-center electron repulsion integrals (ERIs) onto the embedding space is computationally demanding.
To reduce this cost, we employ Gaussian density fitting (GDF)~\citep{whitten1973columbic,sun2017gaussian} to transform the four-center ERIs into three-center ERIs before projection, which substantially lowers computational overhead. Additional implementation details are given in Ref.~\citenum{cui2020efficient}.

To enforce the correct electron distribution on the impurity, a chemical potential $\mu_\text{imp}$ is applied to the impurity, which redistributes electrons between the impurity and the bath. Finite-temperature simulations introduce particle-number fluctuations. We therefore use the trace of the projected mean-field 1RDM as the target electron number of the embedding space, and introduce an additional chemical potential $\mu_\text{gc}$ acting on the entire embedding space, where the subscript ``gc'' denotes the grand canonical ensemble. The effective embedding Hamiltonian thus becomes 
\eqsplit{
H_{\text{emb}}\ \leftarrow\  {H}_{\text{emb}} - \mu_\text{imp}\sum_{p\in \text{imp}} c^{\dag}_pc_p - \mu_\text{gc} \sum_{r\in \text{emb}} c^{\dag}_rc_r.
}

The roles of $\mu_\text{imp}$ and $\mu_\text{gc}$ are illustrated in FIG.~\ref{fig:two_mu_values}. In practice, $\mu_\text{gc}$ is chosen first to ensure the correct total electron number in the embedding space, followed by optimization of $\mu_\text{imp}$ to redistribute electrons between the impurity and bath. Efficient initial guesses for these chemical potentials are critical, as repeated calls to the impurity solver are required for the optimization procedure. 
In Section~\ref{sec:mu_gc}, we provide low-cost approximations for $\mu_\text{gc}$ and show that these approximations are sufficiently accurate to yield the correct electron number in the embedding space.

\subsubsection{Adjusting the embedding electron number\label{sec:mu_gc}}
A direct approach to enforce the correct electron number in the embedding space is to solve
\eqsplit{
\mu_\text{gc} = \arg\min_{\mu}\Big\vert\langle N_e(\mu, \beta)\rangle - N_e^{0}\Big\vert,
}
where $N_e^{0}$ is the target electron number. Evaluating $\langle N_e(\mu, \beta)\rangle$ requires repeated calls to the impurity solver and is therefore computationally expensive.

Here, we provide two cost-friendly approximations for $\mu_\text{gc}$: (1) the mean-field approximation $\mu_\text{gc}^{\text{mf}}$ and (2) the mid-gap approximation $\mu_\text{gc}^{\text{mid}}$. 
The mean-field approximation is obtained by replacing the exact expectation value $\langle N_e(\mu, \beta)\rangle$ with a sum of Fermi-Dirac occupancies:
\eqsplit{
\mu_\text{gc}^{\text{mf}} = \arg\min_{\mu} \Big\vert\sum_i f(\varepsilon_i;  \mu, \beta) - N_e^0 \Big\vert,
}
where $f(\varepsilon_i;  \mu, \beta)$ is the Fermi-Dirac occupancy of the $i$-th MO with orbital energy $\varepsilon_i$.

The mid-gap chemical potential follows from the thermodynamic relation $\mu = (\partial \mathcal{F}/\partial N_e)_{T, V}$,  which leads to the finite-difference approximation 
\eqsplit{
\mu_\text{gc} \approx (\mathcal{F}(N_e^0+1) - \mathcal{F}(N_e^0-1))/2,
}
where $\mathcal{F} = -\ln Z/\beta$ is the canonical Helmholtz free energy.
In the low-temperature limit, $\mathcal{F}(N_e)$ is close to the ground-state energy of the $N_e$-electron sector, $E_0(N_e)$, yielding
\eqsplit{
\mu_\text{gc}^{\text{mid}} \approx [E_0(N_e^0+1) - E_0(N_e^0-1)]/2.
} 

For gapped systems at low temperature, both approximations can provide reasonable estimates of $\mu_\text{gc}$.
We assess the accuracy of these estimates at different values of $\beta$ using $\mu_\text{gc}^{\text{mf}}$ and $\mu_\text{gc}^{\text{mid}}$ in FIG.~\ref{fig:mu_guess_nelectron}. Two systems are considered:
(1) a one-dimensional six-site half-filled Hubbard model with PBC, and (2) a (4e, 6o) embedding Hamiltonian derived from a hydrogen chain with PBC, constructed using two hydrogen atoms in the impurity and a 3-21G basis set. 
Interatomic distances $R = 1.5\bohr$ and $3.0\bohr$ are considered for the hydrogen chain.
Details of these two systems are provided in Appendix~\ref{sec:apdx_ftsolvers}.
Although the absolute energy units differ, i.e., hartree ($E_\text{h}$) for the embedding Hamiltonian and the hopping amplitude $t$ for the Hubbard model, the inverse temperature $\beta$ is defined consistently as the inverse of the corresponding energy unit, which allows direct comparison of numerical $\beta$ values across the two models.

The thermal-average electron number $N_e(\mu)$ is evaluated using finite-temperature full configuration interaction (FT-FCI), and the relative error is computed as $(N_e(\mu) - N_e^0)/N_e^0$.
For the embedding Hamiltonian, $\mu_\text{gc}^{\text{mf}}$ yields small errors across the full temperature range, whereas $\mu_\text{gc}^{\text{mid}}$ shows noticeable deviations at high temperatures ($\beta < 10$). In contrast, for the Hubbard model, both estimates remain accurate for all $\beta$ values considered.

To rationalize this difference, we examine the charge gap,
\eqsplit{
\Delta_c = E_0(N_e^0+1) + E_0(N_e^0-1) - 2E_0(N_e^0).
}

For the embedding Hamiltonian, $\Delta_c  = 1.20 \hartree$  for $R=1.5\bohr$ and $0.38 \hartree$ for  $R=3.0\bohr$, while for the Hubbard model, the gap ranges from $2.09t$ ($U=2t)$ to $8.93t$ ($U=12t$). A larger charge gap permits a wider range of chemical potentials that yield the correct electron number, resulting in reduced numerical sensitivity.

Based on these results, we employ the mean-field chemical potential $\mu_\text{gc}^{\text{mf}}$ in all subsequent simulations, as it is robust across temperatures and computationally inexpensive to evaluate.

\subsubsection{Impurity solvers\label{sec:impurity_solvers}}
The embedding space is an open quantum system embedded in the full lattice, and is therefore naturally described within the grand canonical ensemble. For grand-canonical simulations, we employ FT-FCI based on grand-canonical expansion and finite-temperature density matrix renormalization group (DMRG) based on imaginary-time evolution of a purified matrix product state (MPS)~\cite{feiguin2005ftdmrg}, referred to as pFT-DMRG. 
Since many quantum phenomena of interest occur in the low-temperature regime, we also adopt a simplified low-temperature FCI (LT-FCI) and DMRG (LT-DMRG) approach, in which truncations are introduced for both particle-number fluctuations and the energy-spectrum expansion.
The above solvers are implemented using functionalities from the \textsc{PySCF} and \textsc{Block2} packages~\cite{sun2017pyscf, sun2020pyscf, Zhai2023block2}. Additional technical details are provided in Appendix~\ref{sec:apdx_ftsolvers}.

Below we summarize the low-temperature truncation scheme used in the grand-canonical summation. At low temperature, particle-number fluctuations are strongly suppressed by the charge gap, such that only electron-number sectors close to the target value contribute appreciably. We retain a restricted set of sectors according to the criterion
\eqsplit{\label{eq:lt_truncation1}
&\beta \Delta_c(k) - \mu k \leq \eta_1, 
}
where $\Delta_c(k) = E_{N_e^0 + k}^0 - E_{N_e^0}^0$ is the $k$-particle charge gap, and $k$ denotes the deviation from the target electron number.

Within each electron-number sector, only low-lying many-body eigenstates contribute sufficiently to the thermal average. These states are selected according to
\eqsplit{\label{eq:lt_truncation2}
\beta\,\big(E_{N_e}^i - E_{N_e}^0\big) \leq \eta_2,
}
where $E_{N_e}^0$ is the ground-state energy of the sector with $N_e$ electrons, and $E_{N_e}^i$ is the energy for the $i$th excited state.

The thresholds $\eta_1$ and $\eta_2$ control the relevant particle-number and energy fluctuations at finite temperature. In practice, values in the range $5$-$10$ yield good accuracy for the embedding Hamiltonian. Further discussion and numerical benchmarks are provided in Appendix~\ref{sec:apdx_lt_approx}.

\subsection{DMET self-consistency\label{sec:ftdmet_loop}}

The FT-DMET algorithm follows the same overall structure as the ground-state DMET (GS-DMET) approach. First, a mean-field solution of the full lattice is obtained in the presence of a correlation potential $V_c$, from which the embedding space is constructed. A higher-level solver is then applied to the embedding Hamiltonian, and $V_c$ is optimized by matching the 1RDMs from the mean-field and embedding solutions,
\eqsplit{\label{eq:cost_func}
V_c = \arg\min_{v} \left(D_\text{HF}^\text{imp}(v) - D_\text{emb}^\text{imp}\right)^2,
}
where the superscript ``imp'' indicates that only the impurity blocks of the 1RDMs are matched. 
A flow chart of the DMET algorithm is provided in Appendix~\ref{sec:apdx_dmet_alg}.

At finite temperature, several modifications are introduced:
\begin{enumerate}
\item The mean-field solution is evaluated at finite temperature.
\item An extended bath is required in the medium- to high-temperature regime.
\item The embedding Hamiltonian is solved at finite temperature.
\item An analytical finite-temperature gradient of Eq.~\eqref{eq:cost_func} is employed~\cite{cui2020threeband}.

\end{enumerate}

At low temperature, the finite-temperature correlation potential is expected to be close to its ground-state counterpart. To reduce the computational cost, a one-shot FT-DMET approach can be adopted: a converged $V_c$ and impurity chemical potential $\mu_\text{imp}$ are first obtained from GS-DMET, followed by a single FT-DMET iteration using these ground-state values as the initial guess.

\begin{figure}
    \centering
    \includegraphics[width=0.85\linewidth]{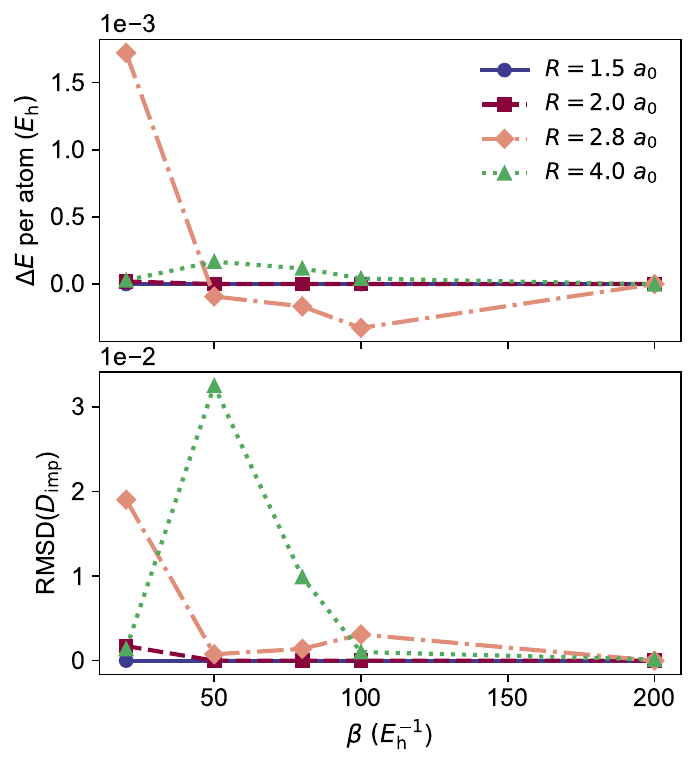}
    \caption{One-shot FT-DMET errors for a hydrogen chain, referenced to converged FT-DMET results.}
    \label{fig:oneshot_error}
\end{figure}

We assess this strategy by simulating a hydrogen chain with a two-atom unit cell and $5$ $k$-points using the STO-6G basis set. FCI and FT-FCI are employed as the ground-state and finite-temperature impurity solvers, respectively, to eliminate solver-related errors.
FIG.~\ref{fig:oneshot_error} shows the error of one-shot FT-DMET relative to fully self-consistent FT-DMET for various interatomic distances and inverse temperatures. 
Both the energy per atom error and the root-mean-square deviation (RMSD) of the impurity 1RDM remain small across the tested range ($\beta = 20$-$200 \hartree^{-1}$), though errors increase with temperature. The larger errors observed at $R = 4.0\bohr$ and $2.8\bohr$ arise because the system exhibits antiferromagnetic (AFM) order at large $R$, where spin correlations are particularly sensitive to temperature.

{
\subsection{Implementation}
The \textit{ab initio} FT-DMET code is implemented within the \textsc{LibDMET} framework~\cite{cui2020efficient}, a ground-state DMET package for \textit{ab initio} systems. Core quantum chemistry functionalities, including crystalline system initialization, $k$-point sampling, and Hartree--Fock self-consistent field (SCF) calculations, are provided by the \textsc{PySCF} package~\cite{sun2017pyscf,sun2020pyscf}. The FT-FCI and LT-FCI solvers are built upon the FCI infrastructure in \textsc{PySCF}, while the pFT-DMRG and LT-DMRG solvers are implemented using functionalities from the \textsc{Block2} package~\cite{Zhai2023block2}.}

\section{Applications}\label{sec:result}

We apply FT-DMET to investigate finite-temperature phases of one- and two-dimensional hydrogen lattices. Throughout this section, results are presented as functions of the temperature $T = 1/\beta$, with the Boltzmann constant set to $k_\mathrm{B}=1$. 

For the results in Sections~\ref{sec:result_mag} and \ref{sec:result_docc}, we consider a hydrogen chain with two atoms per unit cell and an $11$-point $k$-mesh. This setup enforces equivalence of local observables on the two sites (up to a sign change for staggered quantities), ensuring that the magnitudes of local observables are uniform within the unit cell and not influenced by explicit dimerization. A 3-21G basis set is used unless otherwise noted. The impurity problem is solved using FT-FCI.

For the dimerization study in Section~\ref{sec:dimer}, we use a hydrogen chain {with a bond length of $2.8\ a_0$,} a four-atom impurity, a $9$-point $k$-mesh, and the STO-6G basis set. The impurity problem is solved using LT-DMRG with a bond dimension of $400$ and truncation thresholds $\eta_1=\eta_2=8$. Finally, in Section~\ref{sec:2d_hlatt}, we study a two-dimensional hydrogen square lattice using a $2\times2$ impurity cluster and a $3\times3$ $k$-point mesh. The STO-6G basis set is used, and the impurity problem is solved with LT-DMRG using a bond dimension of $600$ and truncation thresholds $\eta_1=\eta_2=8$.

\begin{figure}
    \centering
    \begin{subcaptionbox}{Magnetic moment with 3-21G basis set \label{sub:mag_321g}}{\includegraphics[width=0.8\linewidth]{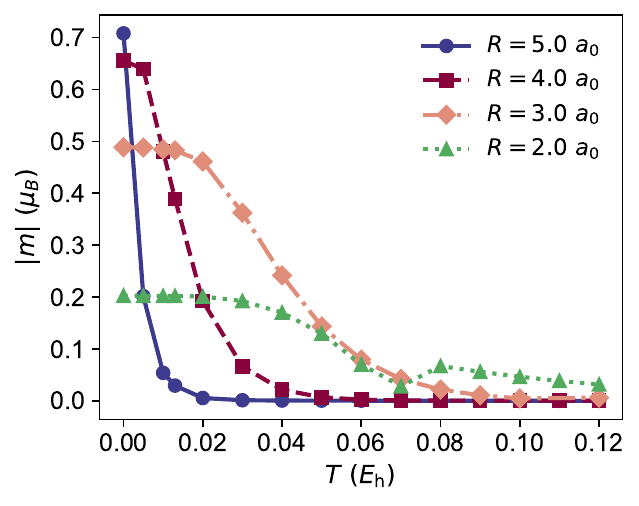}}
    \end{subcaptionbox}
    \vspace{0.5em}
    
    \centering
    \begin{subcaptionbox}{Comparing STO-6G and 3-21G \label{sub:mag_sto6g_vs_321g}}{\includegraphics[width=0.8\linewidth]{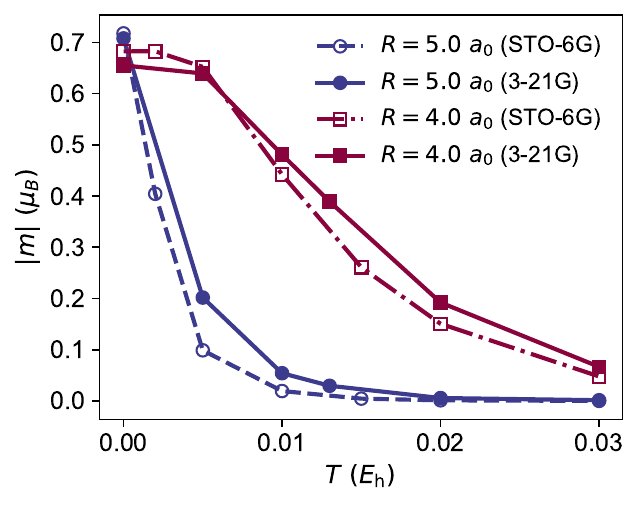}}
    \end{subcaptionbox}
    \caption{Averaged absolute magnetic moment of the hydrogen chain as a function of temperature.
(a) Results at several interatomic distances computed using the 3-21G basis set.
(b) Results at large interatomic distances comparing the STO-6G and 3-21G basis sets. }
    \label{fig:result_magnetic}
\end{figure}

\subsection{Hydrogen chain \label{sec:result_hchain}}
{The hydrogen chain provides a simple one-dimensional crystalline system that is often compared with the one-dimensional Hubbard model, but exhibits richer physics. Unlike the Hubbard model, the hydrogen lattice features long-range Coulomb interactions and multiband effects arising from the use of non-minimal basis sets. These distinctions lead to qualitatively different ground-state and finite-temperature behavior, making the hydrogen chain a popular benchmark system for \textit{ab initio} many-body simulations~\cite{motta2017hchain, motta2020hchain, liu2020ftafqmc}.}

\subsubsection{Magnetic moment\label{sec:result_mag}}
The local magnetic moment on atom $i$ is defined as
\eqsplit{
m_i = n_{i\uparrow} - n_{i\downarrow},
}
and is reported in units of the Bohr magneton $\mu_\text{B}$. The spin-resolved occupancies $n_{i\sigma}$ are obtained by summing the diagonal elements of the spin-resolved 1RDM over all IAOs assigned to atom $i$. A staggered pattern of nonzero $m_i$ with alternating signs indicates antiferromagnetic (AFM) order, while $m_i = 0$ corresponds to a paramagnetic (PM) state in the absence of explicit symmetry breaking.

FIG.~\ref{fig:result_magnetic} shows the absolute magnetic moment averaged over hydrogen atoms,
\eqsplit{
|m| = \frac{1}{\nimp} \sum_{i=0}^{\nimp-1} |m_i|.
}
Results obtained with the 3-21G basis set are shown in FIG.~\ref{sub:mag_321g}, which includes both $1s$ and split-valence $s$-type functions. At the ground state, $|m|$ increases monotonically with $R$, reflecting progressive electron localization and formation of local moments as the chain approaches the atomic limit. The temperature dependence of $|m|$ is strongly $R$-dependent. At $R = 5 \bohr$, weak orbital overlap places the system deep in a strong-correlation regime, effectively described by a Hubbard model with large $U/t$, consistent with the Mermin-Wagner theorem~\cite{mermin1966absence}. As $R$ decreases, increased orbital overlap enhances itinerancy and spin exchange, producing a broader and smoother suppression of AFM order with temperature. The kink at $R = 2.0 \bohr$ and $T \approx 0.07 \hartree$ corresponds to a spin sign flip on each site, reflecting thermal averaging over symmetry-related AFM states rather than a physical phase transition.

FIG.~\ref{sub:mag_sto6g_vs_321g} compares the temperature dependence of $|m|$ obtained with the STO-6G and 3-21G basis sets to assess multi-band effects. The faster decay of $|m|$ in the minimal STO-6G basis reflects reduced variational flexibility, which enhances localization and sharpens the AFM-PM crossover. In contrast, the additional radial flexibility in 3-21G slightly increases charge fluctuations, moderating the temperature dependence of the magnetic moment. Despite these quantitative differences, both basis sets reproduce the same qualitative trends, indicating that the minimal basis is sufficient when using IAOs.

\begin{figure}
    \centering
\includegraphics[width=0.8\linewidth]{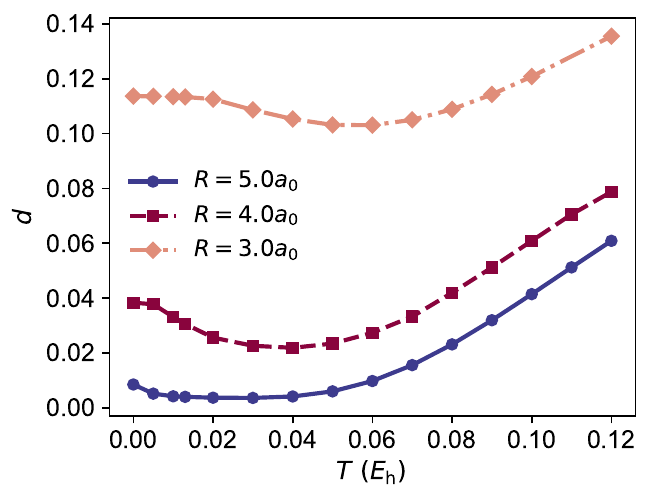}
    \caption{Double occupancy of hydrogen chain as a function of temperature.}
    \label{fig:result_docc}
\end{figure}
\subsubsection{Double occupancy\label{sec:result_docc}}
The double occupancy on atom $i$ is defined as
\eqsplit{
d_i = \langle n_{i\uparrow} n_{i\downarrow} \rangle,
}
which measures the probability that two opposite-spin electrons occupy the same atomic site. As a local observable, $d_i$ directly reflects the strength of electronic correlations: larger values indicate enhanced charge fluctuations and more itinerant electronic behavior, while reduced values signal strong correlations and local-moment formation. In the hydrogen chain, the ground-state $d$ decreases with increasing interatomic distance $R$, consistent with a crossover from a band-like regime at small $R$ to a Mott-like regime with well-formed local moments at large $R$.

FIG.~\ref{fig:result_docc} shows the site-averaged double occupancy as a function of temperature for several interatomic distances larger than equilibrium, using the same computational setup as in FIG.~\ref{sub:mag_321g}. For each $R$, a shallow minimum in $d(T)$ is observed. The temperature at which this minimum occurs closely coincides with the rapid decrease of the magnetic moment in FIG.~\ref{sub:mag_321g}.

The non-monotonic temperature dependence of $d$ reflects the competition between AFM correlations and charge fluctuations. As temperature increases from the ground state, suppression of AFM correlations reduces virtual hopping processes that contribute to residual charge fluctuations, causing $d$ to decrease and reach a minimum. At higher temperature, thermal activation of real charge excitations dominates, leading to an increase in $d$. This minimum therefore marks a crossover from an exchange-dominated regime at low temperature to a thermally activated charge-fluctuation regime at higher temperature.
This behavior is also referred to as the Pomeranchuk effect~\cite{richardson1997pomeranchuck}, where the system ``orders'' its spatial configuration to accommodate spin disorder.
\begin{figure*}[t!]
    \centering
    \begin{subcaptionbox}{Bond order difference \label{sub:dimer_imp4}}{\includegraphics[width=0.325\linewidth]{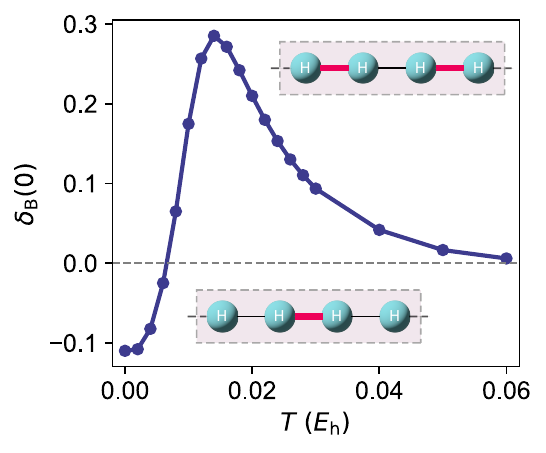}}
    \end{subcaptionbox}
    \begin{subcaptionbox}{Spin-spin correlation \label{sub:spin_corr}}{\includegraphics[width=0.325\linewidth]{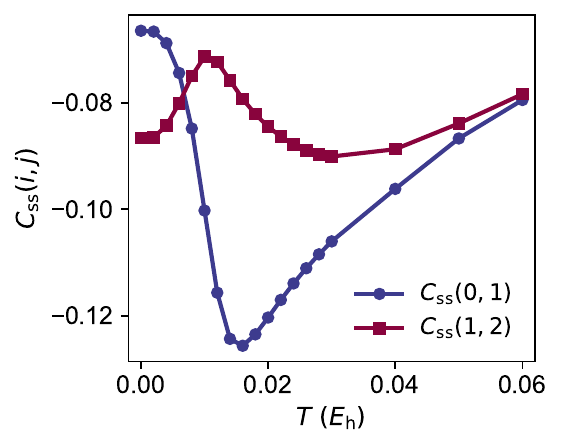}}
    \end{subcaptionbox}
    \begin{subcaptionbox}{Magnetic moment \label{sub:mag_imp4}}{\includegraphics[width=0.325\linewidth]{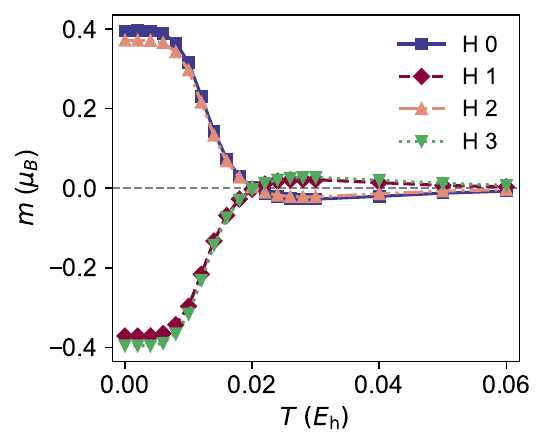}}
    \end{subcaptionbox}
    \caption{Dimerization of the hydrogen chain {at $R = 2.8\ a_0$}. (a) Bond-order difference, (b) spin-spin correlation, and (c) magnetic moment as functions of temperature. Bold red lines in (a) indicate stronger bonds within the unit cells.}
    \label{fig:dimerization}
\end{figure*}

    

\subsubsection{Dimerization\label{sec:dimer}}

The hydrogen chain at large interatomic separation $R$ provides a prototype for dimerization, often described as a manifestation of Peierls’ theorem~\cite{peierls1955quantum, motta2020hchain}. We probe the stability of this structural distortion at finite temperature via the difference between adjacent bond orders:
\eqsplit{
\delta_\text{B}(i) &= B_{i, i+1} - B_{i+1, i+2},
}
where $B_{i, j} = D_{i, j} + D_{j, i}$ is the symmetric hopping amplitude derived from the 1RDM.

FIG.~\ref{sub:dimer_imp4} shows the temperature dependence of $\delta_\text{B}(0)$. At the ground state, the bond order is non-uniform: $B_{1,2} > B_{0,1}$, reflecting a stronger singlet-like correlation between the central atoms. This inhomogeneity arises from the embedding environment: central atoms form a stable dimer, while edge atoms (sites 0 and 3) are more strongly entangled with the bath, weakening their internal bond order. 

With increasing temperature, long-range AFM order is suppressed, accompanied by a redistribution of electronic correlations. The sign change of $\delta_\text{B}(0)$, a bond-order swap, signals structural reorganization driven by competition between magnetic and lattice entropy. 
This behavior is consistent with a Pomeranchuk-like mechanism, in which the suppression of the AFM moment increases spin entropy that stabilizes an alternative ordered phase at finite temperature.
At higher temperatures ($T > 0.02 \hartree$), thermal fluctuations dominate, melting the dimerized phase.

The interplay between spin correlations and structural order is clarified via nearest-neighbor spin-spin correlation functions:
\eqsplit{
C_\text{ss}(i, j) = \langle S^z_i S_j^z\rangle - \langle S_i^z\rangle \langle S_j^z\rangle, \quad S^z_i = \frac{n_{i\uparrow} - n_{i\downarrow}}{2}.
}
Negative values indicate AFM coupling, and the magnitude $|C_\text{ss}(i,j)|$ measures local singlet character. As shown in FIG.~\ref{sub:spin_corr}, $C_\text{ss}(i,j)$ closely tracks the dimerization patterns: at low temperature, $C_\text{ss}(1,2)$ is stronger than $C_\text{ss}(0,1) = C_\text{ss}(2,3)$, and the crossover in correlations coincides with the inversion of $\delta_\text{B}(0)$, indicating that the bond-order swap responds to spin redistribution.

The local magnetic moment $m$ for each impurity atom in FIG.~\ref{sub:mag_imp4} further confirms this behavior. The crossover in $\delta_\text{B}(0)$ coincides with a sharp decrease in staggered magnetization. The peak in $\delta_\text{B}(0)$ at $T \approx 0.015 \hartree$ occurs as AFM order vanishes, suggesting that dimerization compensates for diminishing spin correlations. Compared with the 2-atom impurity (FIG.~\ref{fig:result_magnetic}), AFM order decays earlier in the 4-atom cluster, showing that larger impurities better capture collective correlations and sharpen transitions toward the thermodynamic limit.

\subsection{Two-dimensional hydrogen square lattice\label{sec:2d_hlatt}}
{We extend our study to the two-dimensional hydrogen square lattice, a realistic \textit{ab initio} analogue of the two-dimensional Hubbard model. As discussed in Section~\ref{sec:result_hchain}, compared to the Hubbard model, the hydrogen square lattice exhibits richer physics due to long-range Coulomb interactions and multiband effects. Consequently, it serves as a more realistic prototypical system for complex two-dimensional correlated materials, such as copper-based high-temperature superconductors~\cite{zheng2016ground, zheng2017stripe}.}
Compared to the one-dimensional chain, the higher coordination number ($z=4$) in two dimensions enhances the stability of AFM correlations against thermal fluctuations. This robustness of two-dimensional spin order underpins the design of quantum error-correcting codes such as the surface code~\cite{fowler2012surface}.

From a computational perspective, the two-dimensional geometry poses additional challenges for DMET. The larger boundary ``surface area'' of a two-dimensional impurity cluster relative to its volume increases entanglement with the bath orbitals---following the area law of entanglement entropy---making it more demanding to recover bulk-like behavior without enlarging the impurity.

\begin{figure}[t!]
    \centering
    \includegraphics[width=0.8\linewidth]{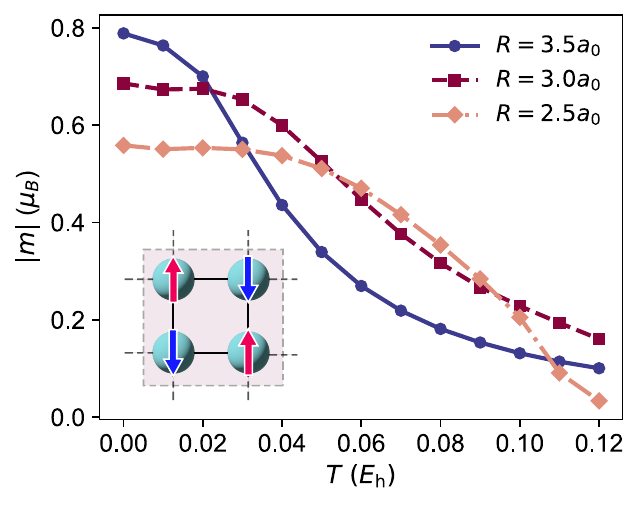}
    \caption{Site-averaged magnetic moment of the two-dimensional hydrogen square lattice.}
    \label{fig:2d_hlatt}
\end{figure}

FIG.~\ref{fig:2d_hlatt} shows the site-averaged magnetic moment of the two-dimensional hydrogen lattice. Consistent with the higher coordination, AFM order is significantly more robust than in one dimension at comparable interatomic distances $R$. The decay of the magnetic moment $m$ with temperature $T$ is slower, reflecting the larger energy barrier for thermal excitations in two dimensions. Among the bond lengths considered ($R = 3.5, 3.0$, and $2.5 \bohr$), AFM order persists to higher temperatures at shorter $R$, owing to the increased electronic hopping amplitude $t$. The temperature dependence qualitatively reproduces the magnetic phase behavior observed in the two-dimensional Hubbard model at half-filling~\cite{sun2020ftdmet}.

\section{Conclusion}\label{sec:conclusion}
In this work, we presented a finite-temperature extension of density matrix embedding theory (FT-DMET) for realistic crystalline systems and developed a practical framework for its \textit{ab initio} implementation. By testing our method on periodic hydrogen chains and square lattices with both minimal and multi-band basis sets, we demonstrate that FT-DMET can handle \textit{ab initio} systems with long-range Coulomb interactions and multiple orbitals per site. Our simulations captured rich finite-temperature phase behavior of the hydrogen lattice, including the Pomeranchuk effect in one dimension and enhanced stability of long-range order in two dimensions.
Although the method is demonstrated using hydrogen lattices, we expect the framework to be applicable to larger and more complex systems with increased computational resources and extensive parallelization of the impurity solvers. For example, with the LT-DMRG solver, embedding spaces of approximately 30 orbitals should be accessible. Furthermore, the development of active-space strategies could further enlarge the embedding sizes that FT-DMET can handle.

\begin{acknowledgments}
This research is supported by a start-up fund from the Rutgers University.
We acknowledge the Office of Advanced Research Computing (OARC) at Rutgers University for computational resources.
C. S. acknowledge Huanchen Zhai and Lingqing Peng for useful discussions.
\end{acknowledgments}

\section*{Data Availability Statement}
The data that support the findings of this study are available from the corresponding author upon reasonable request.

\appendix

\section{Finite-temperature impurity solvers\label{sec:apdx_ftsolvers}}

This appendix summarizes and analyzes the finite-temperature impurity solvers implemented in the package, including finite-temperature full configuration interaction (FT-FCI), purification-based finite-temperature DMRG (pFT-DMRG), and their low-temperature counterparts, LT-FCI and LT-DMRG. All solvers operate in the grand-canonical ensemble.

Two representative Hamiltonians are used for the simulations presented in this appendix:
(1) a (4e,6o) embedding Hamiltonian obtained from DMET using the interacting-bath construction for the hydrogen chain, with two impurity atoms and three $k$ points at $R = 1.5\bohr$ in the STO-6G basis; and
(2) a (6e,6o) half-filled one-dimensional Hubbard model with periodic boundary conditions (PBC) and interaction strength $U = 8t$. The Hamiltonians are given by
\eqsplit{\label{eq:si_two_hams}
H_{\text{emb}} =& \sum_{\alpha=\uparrow, \downarrow}\sum_{ij=0}^{\norb-1} h_{ij,\alpha} a^\dagger_{i\alpha} a_{j\alpha} \\
&+ \sum_{\alpha,\beta=\uparrow, \downarrow}\sum_{ijkl = 0}^{\norb - 1} (ij|kl)_{\alpha,\beta} a^\dagger_{i\alpha} a^\dagger_{k\beta} a_{l\beta} a_{j\alpha}, \\ 
H_{\text{Hub}} =& -t \sum_{\alpha=\uparrow, \downarrow}\sum_{i=0}^{\norb-1} a^\dagger_{i\alpha} a_{(i+1)\alpha} + U\sum_{i=0}^{\norb-1} n_{i\uparrow}n_{i\downarrow},
}
where $t$ denotes the hopping amplitude and $U$ the on-site Coulomb interaction strength in the Hubbard model. PBC impose $i+1=0$ when $i=\norb-1$ in the hopping term of $H_{\text{Hub}}$. A temperature-dependent chemical potential is added to each Hamiltonian to enforce the target electron number,
\eqsplit{
-\mu(\beta) \sum_{\alpha=\uparrow,\downarrow}\sum_{i=0}^{\norb-1} n_{i\alpha}.
}
Energies are reported in units of $\hartree$ for the embedding system and $t$ for the Hubbard model. As discussed in the main text, the embedding Hamiltonian exhibits a smaller charge gap and a denser low-energy spectrum than the Hubbard Hamiltonian, rendering it more sensitive to thermal fluctuations.

\subsection{FT-FCI}

The implementation of FT-FCI in the grand-canonical ensemble is straightforward. Algorithm~\ref{alg:fted} summarizes the procedure used to evaluate thermal averages of the partition function ($Z$), the electronic energy ($E$), and the one-particle reduced density matrix ($D_1$). The inputs are the Hamiltonian $H$, the number of orbitals $\norb$, the inverse temperature $\beta = 1/T$ (with $k_\text{B}=1$), and the chemical potential $\mu$. In all calculations, the same chemical potential is applied to both spin species.

The computational bottleneck of FT-FCI is the \texttt{DiagHam} routine, with the cost peaking at $N_\uparrow = N_\downarrow = \norb/2$. Consequently, FT-FCI can handle system sizes comparable to ground-state FCI, but with a substantial overhead arising from the $(\norb + 1)^2$ loops over particle numbers. If the Hamiltonian conserves spin symmetry, the number of required loops can be reduced to $(\norb + 2)(\norb + 1)/2$.

\begin{algorithm}[t!]
\caption{Grand-canonical FT-FCI algorithm}\label{alg:fted}
\KwIn{$H$, $\norb$, $\beta$, $\mu$}
\KwOut{$E$, $D_1$}
Initialize $Z = 0$, $E = 0$, $D_1 = [0]$ \\
\For{$N_{\uparrow} \gets 0$ \KwTo $\norb$}{
\For{$N_{\downarrow} \gets 0$ \KwTo $\norb$}{
$N_e = N_{\uparrow} + N_{\downarrow}$, \\
$\varepsilon, \mathbf{v} = $ \texttt{DiagHam}[$H$, $N_{\uparrow}$, $N_{\downarrow}$],\\
$Z = Z + $ \texttt{Sum}[exp($-\varepsilon + \mu N_e$)],\\
$E = E+$ \texttt{Sum}[$\varepsilon * \exp(-\varepsilon + \mu N_e$)],\\
\For{$i \gets 1$ \KwTo len($\varepsilon$)}{
$D_1 = D_1 +\exp(-\varepsilon_i + \mu N_e) * $ \texttt{MakeRDM1}[$v_i$, $\norb$, $N_{\uparrow}$, $N_{\downarrow}$]
}
}
}
\Return $E = E/Z$, $D_1 = D_1/Z$
\end{algorithm}

\subsection{pFT-DMRG\label{sec:si_pft_dmrg}}

We adapt the purification-based finite-temperature DMRG (pFT-DMRG) approach~\cite{feiguin2005ftdmrg} to fermionic systems~\cite{sun2020ftdmet, Zhai2023block2} and briefly summarize the algorithm here. This method is also referred to as the ancilla approach. For a given physical system, an identical copy called the ancilla system is introduced. The combined physical and ancilla degrees of freedom define a purified system, represented by a matrix product state (MPS) $|\Psi(\beta)\rangle$.

Starting from the infinite-temperature state $|\Psi(0)\rangle$, the finite-temperature state is obtained by imaginary-time evolution,
\eqsplit{
|\Psi(\beta)\rangle = e^{-\beta H/2}|\Psi(0)\rangle,
}
where the Hamiltonian $H$ acts only on the physical sites. Grand-canonical expectation values are evaluated as
\eqsplit{
\langle A \rangle(\beta) = \frac{\langle \Psi(\beta) | A | \Psi(\beta)\rangle}{\langle \Psi(\beta) | \Psi(\beta)\rangle}.
}

Due to error accumulation and entanglement growth during imaginary-time evolution, this method is generally effective at higher temperatures but can incur significant errors and computational cost at low temperatures. To reduce finite-step errors, we employ a fourth-order Runge-Kutta (RK4) scheme for the imaginary-time evolution. Let $\tau$ denote the time-step size. For a symmetrized Trotter-Suzuki decomposition, the local truncation error scales as $\mathcal{O}(\tau^3)$, whereas for RK4 it scales as $\mathcal{O}(\tau^5)$. This higher-order convergence permits the use of larger time steps in our simulations. We also find that excessively small $\tau$ values are not always advantageous, as they require more time steps and thus provide more opportunities for error accumulation during the evolution.

\subsection{LT-FCI and LT-DMRG}

We implement the low-temperature FCI and DMRG solvers following Eqs.~\eqref{eq:si_lt_truncation1} and \eqref{eq:si_lt_truncation2} in the next section. To obtain the low-energy states within each electron-number sector, we adopt two strategies:
(1) simultaneous computation of the lowest $L_r$ states using block diagonalization methods, such as the Davidson algorithm, and
(2) sequential computation of the $l$-th state by projecting out the previously obtained $l-1$ states.
The block Davidson approach is generally more stable, whereas the sequential projection strategy is less memory-intensive. The latter is therefore preferred in DMRG-based solvers, where memory usage is often the dominant computational bottleneck.



\begin{figure}[t!]
\centering
\includegraphics[width=\linewidth]{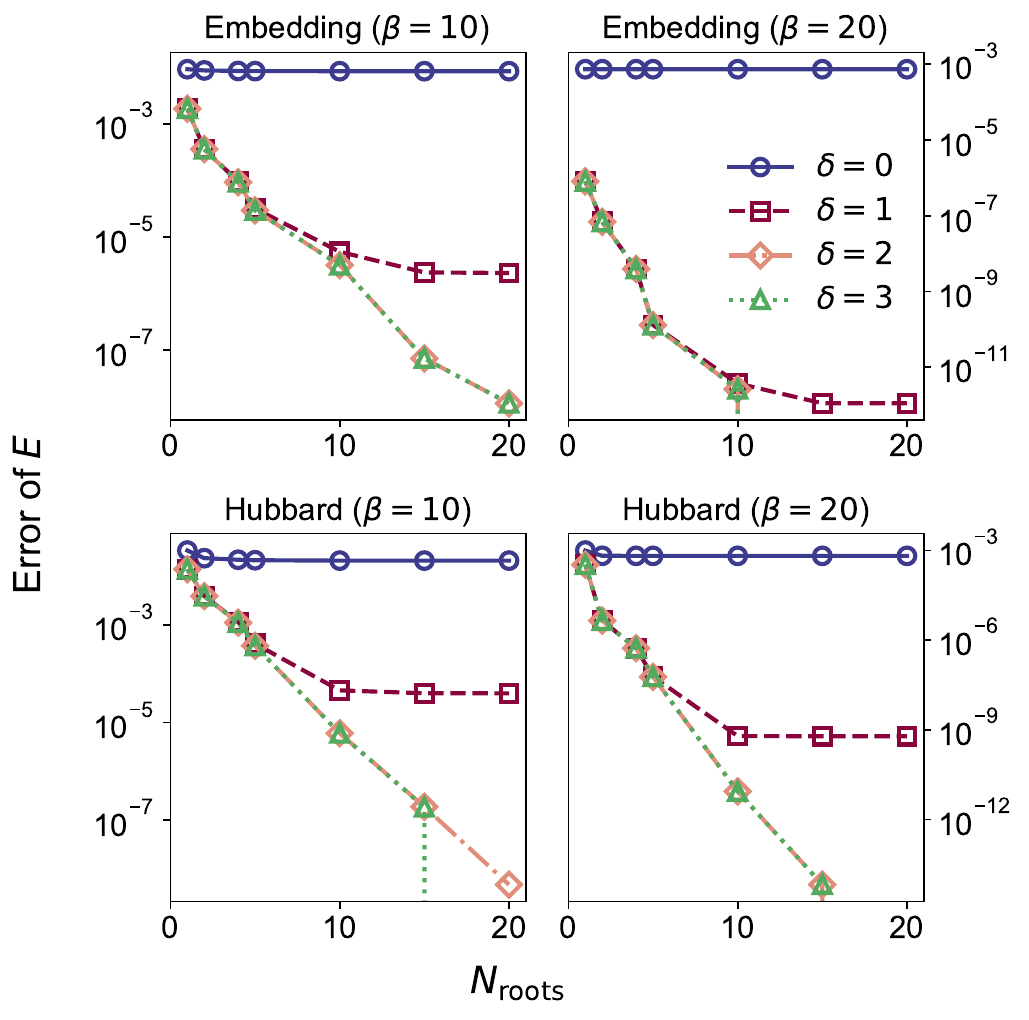}
\caption{Energy error of the truncated grand-canonical summation in Eq.~\eqref{eq:si_truncated_sum} for the embedding system and the one-dimensional Hubbard model. }
\label{fig:si_ltfci_energy}
\end{figure}





\begin{figure}[t!]
\centering

\begin{subcaptionbox}{Embedding Hamiltonian\label{sub:si_cgap_embham}}{\includegraphics[width=\linewidth]{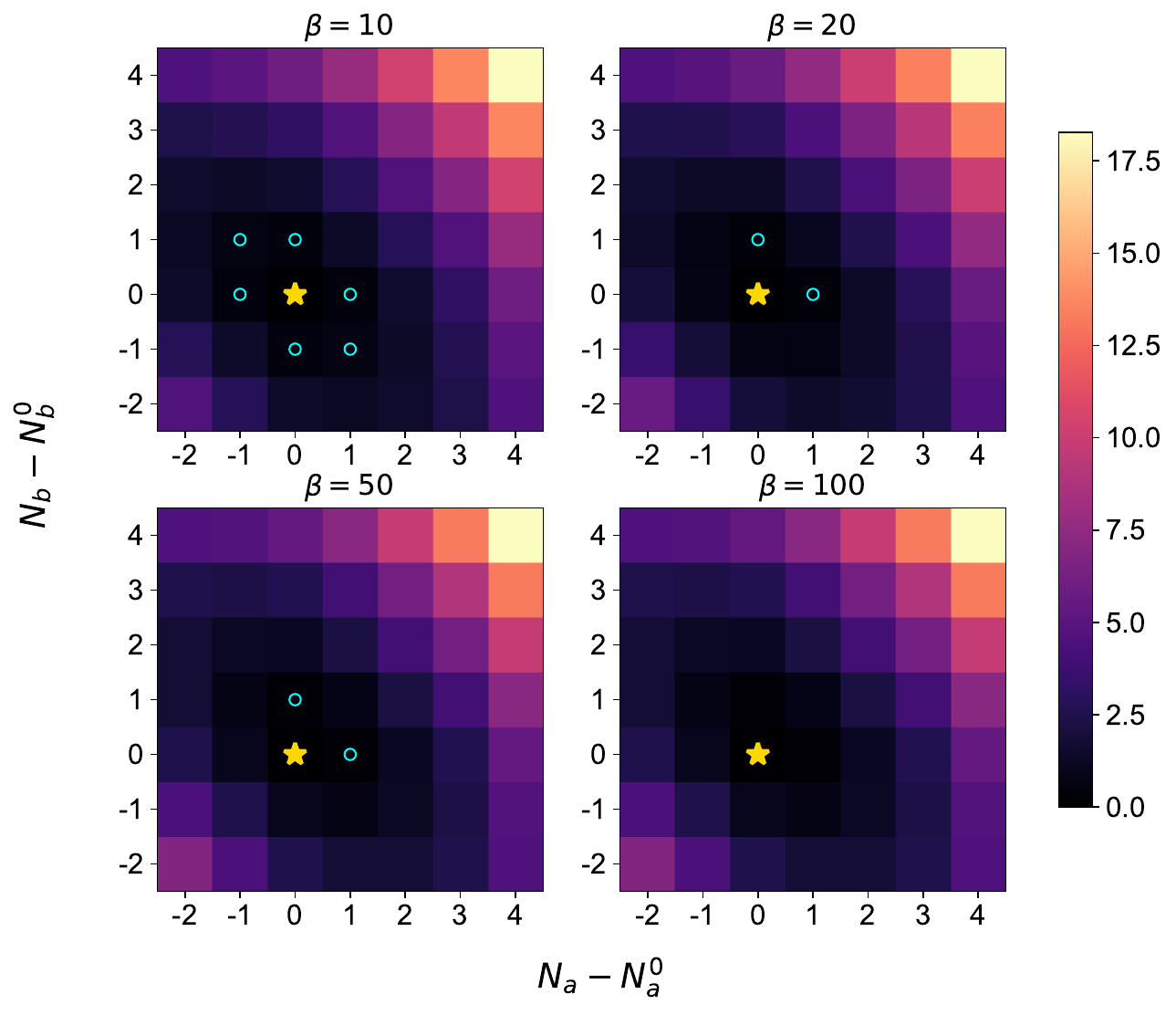}}
\end{subcaptionbox}

\vspace{0.8em}

\begin{subcaptionbox}{Hubbard Hamiltonian\label{sub:si_cgap_hub}}{\includegraphics[width=\linewidth]{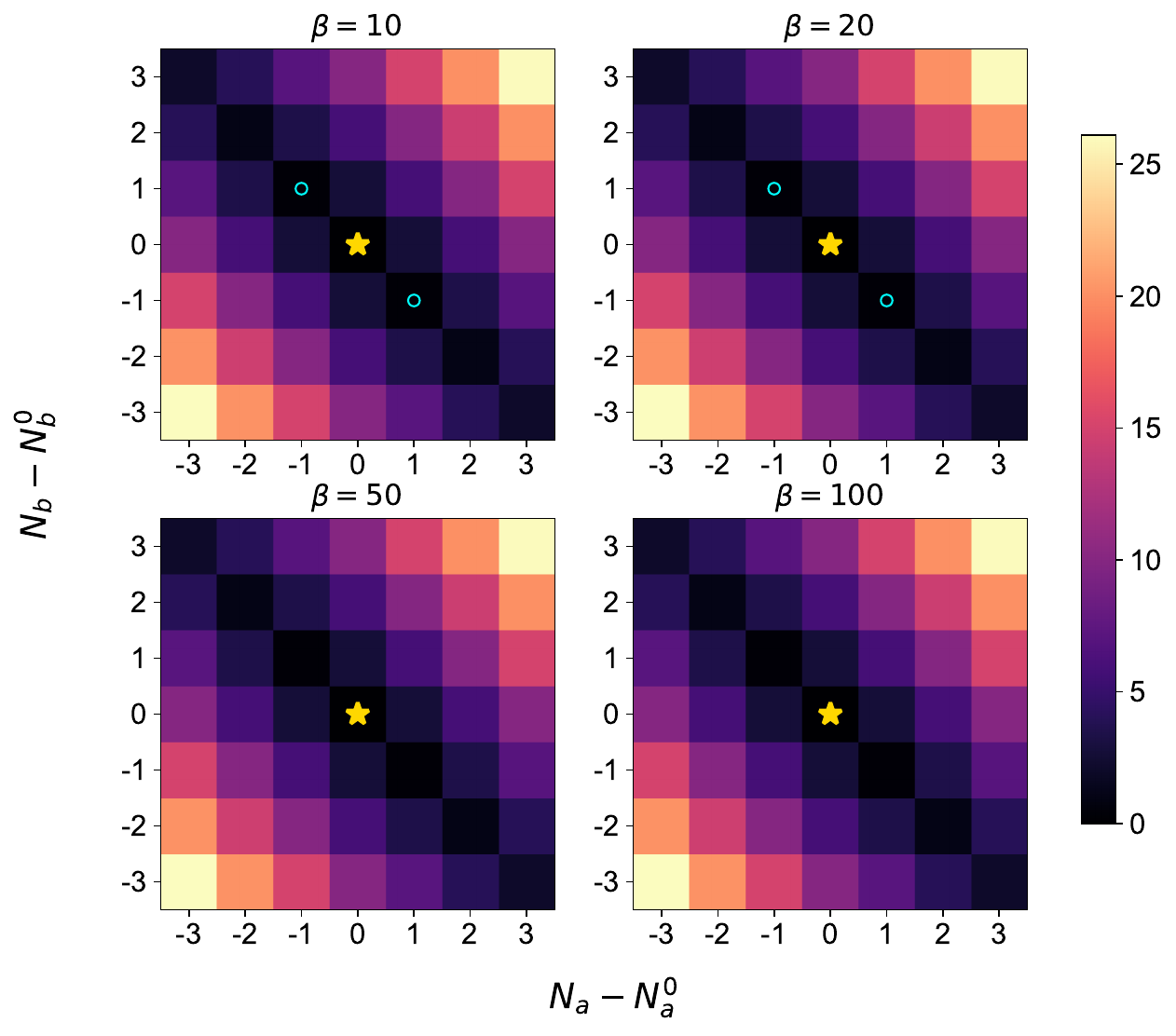}}
\end{subcaptionbox}

\caption{Charge-gap heatmaps for different electron-number sectors for (a) the embedding system and (b) the one-dimensional Hubbard model. The star symbols indicate the target electron numbers, while circles mark the sectors included according to Eq.~\eqref{eq:si_lt_truncation1} with $\eta_1 = 10$.
}
\label{fig:si_charge_gap_heatmap}
\end{figure}

\begin{figure}
    \centering
    \includegraphics[width=1
    \linewidth]{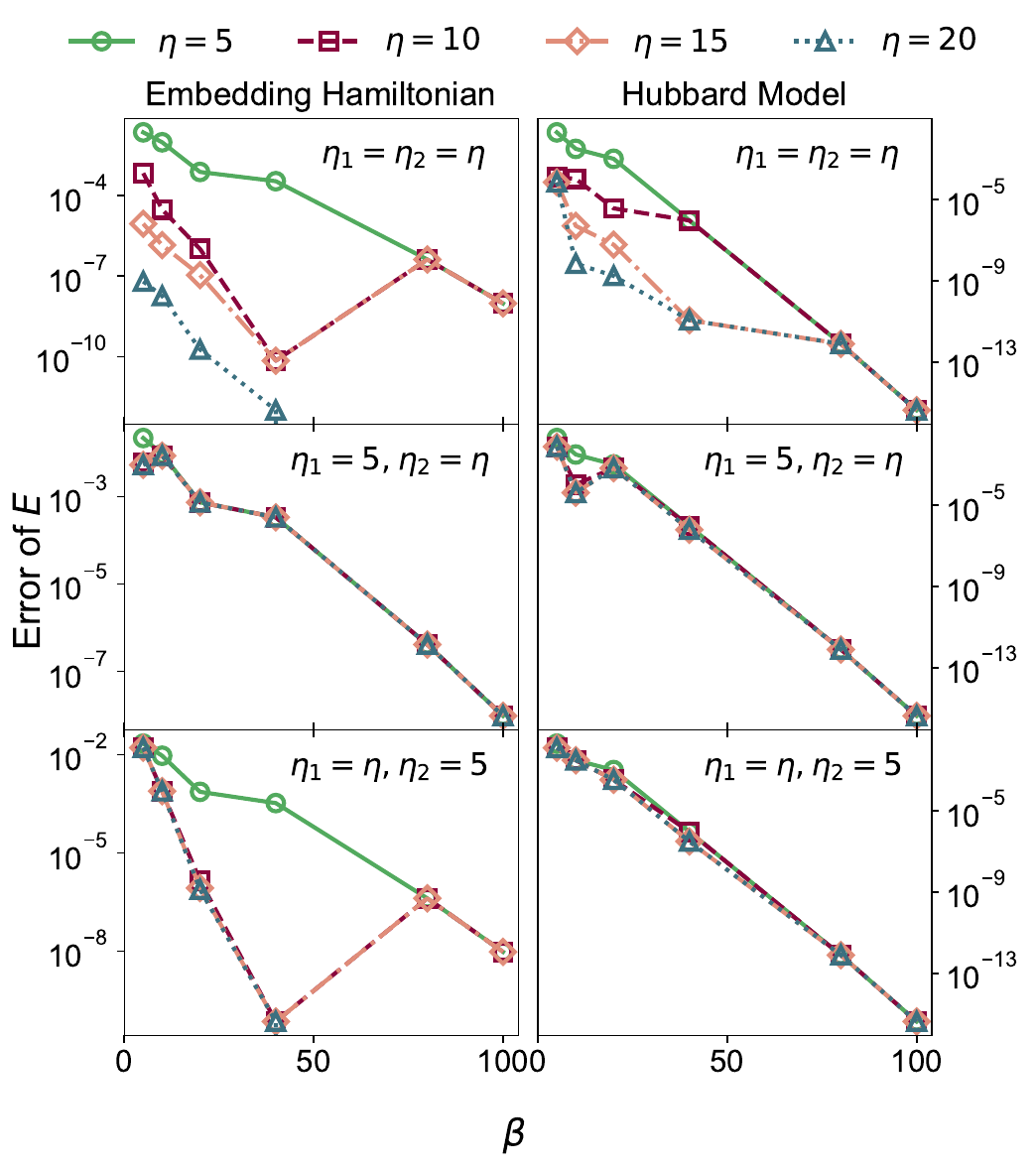}
    \caption{Energy error of the truncated grand-canonical summation using the criteria in Eqs.~\eqref{eq:si_lt_truncation1} and \eqref{eq:si_lt_truncation2}. The parameter $\eta_1$ controls truncation of electron-number fluctuations, while $\eta_2$ controls truncation of excited states.
    }
    \label{fig:si_trunc_eta}
\end{figure}

\section{Low-temperature truncation\label{sec:apdx_lt_approx}}
At low temperature, electron-number and energy fluctuations are small, allowing the full grand-canonical expansion to be approximated by a truncated one:
\eqsplit{\label{eq:si_truncated_sum}
\sum_{N_{\uparrow}=0}^{\norb}\sum_{N_{\downarrow}=0}^{\norb} \sum_{i=1}^{M_{N_e}} 
\;\; \longrightarrow\;\;
\sum_{N_{\uparrow}=N_{\uparrow}^0 - \delta}^{N_{\uparrow}^0 + \delta} \sum_{N_{\downarrow}=N_{\downarrow}^0 - \delta}^{N_{\downarrow}^0 + \delta} \sum_{i=1}^{m_{N_e}}
}
where $M_{N_e}$ is the Hilbert-space dimension of the $(N_{\uparrow},N_{\downarrow})$ sector for a system with $\norb$ orbitals. In the truncated expansion, only sectors near the target electron numbers $(N_{\uparrow}^0,N_{\downarrow}^0)$ are retained, and within each sector only the lowest $m_{N_e} \ll M_{N_e}$ eigenstates are included.

We next examine the choice of $\delta$ and $m_{N_e}$ for the two systems defined in Eq.~\eqref{eq:si_two_hams}, using FCI to solve each electron-number sector. For simplicity, we fix $m_{N_e}$ to be the same for all $N_e$. The resulting total-energy error relative to the full grand-canonical expansion is shown in Fig.~\ref{fig:si_ltfci_energy}. Since $\delta$ is applied independently to both spin sectors, $\delta = 1$ corresponds to sampling nine particle-number sectors: $\{N_{\uparrow}^0-1, N_{\uparrow}^0, N_{\uparrow}^0+1\}\otimes   \{N_{\downarrow}^0-1, N_{\downarrow}^0, N_{\downarrow}^0+1\}$.
For small $\delta$, too few low-energy states contribute to the thermal sum, whereas larger particle-number fluctuations require more low-energy states to be retained. In particular, the canonical ensemble ($\delta = 0$) provides a poor approximation even at low temperature, indicating that particle-number fluctuations are essential for controlling the truncation error. Comparing the two Hamiltonians, electron-number fluctuations play a more prominent role for the embedding Hamiltonian, owing to its smaller charge gap and denser low-energy spectrum relative to the Hubbard model.

Based on these observations, we propose two practical truncation criteria for low-temperature grand-canonical expansions. First, we determine which electron-number sectors to include. We define the $k$-particle charge gap as
\eqsplit{
\Delta_c(k) &= E_{N_e^0 + k}^0 - E_{N_e^0}^0,
}
where $k$ can be positive or negative. We include a sector $k$ only if
\eqsplit{\label{eq:si_lt_truncation1}
\beta\,\Delta_c(k) - \mu k \leq \eta_1.
}

Next, for each electron number $N_e$, we determine the low-energy state truncation according to
\eqsplit{\label{eq:si_lt_truncation2}
\beta\,\big(E_{N_e}^i - E_{N_e}^0\big) \leq \eta_2,
}
where $\eta_1, \eta_2 \geq 1$ are predefined exponent-based truncation thresholds.

The charge-gap distributions for the two systems are shown in Fig.~\ref{fig:si_charge_gap_heatmap}. The target electron number is marked by a star symbol, and the values of $k$ satisfying Eq.~\eqref{eq:si_lt_truncation1} with $\eta_1 = 10$ are highlighted by circles. These charge-gap patterns are consistent with the energy-error trends observed in Fig.~\ref{fig:si_ltfci_energy}. In Fig.~\ref{fig:si_trunc_eta}, we benchmark the total-energy error as a function of $\eta_1$ and $\eta_2$ for three cases: varying $\eta_1 = \eta_2 = \eta$ simultaneously, fixing $\eta_1 = 5$ while varying $\eta_2$, and fixing $\eta_2 = 5$ while varying $\eta_1$. The non-monotonic behavior observed for certain $(\eta_1,\eta_2)$ combinations arises from overly aggressive truncation. For low temperatures, choosing $\eta$ in the range of $5$-$10$ provides a good balance between accuracy and computational cost.


\section{Diagram for DMET algorithm\label{sec:apdx_dmet_alg}}
\begin{figure}[h!]
    \centering
    \includegraphics[width=\linewidth]{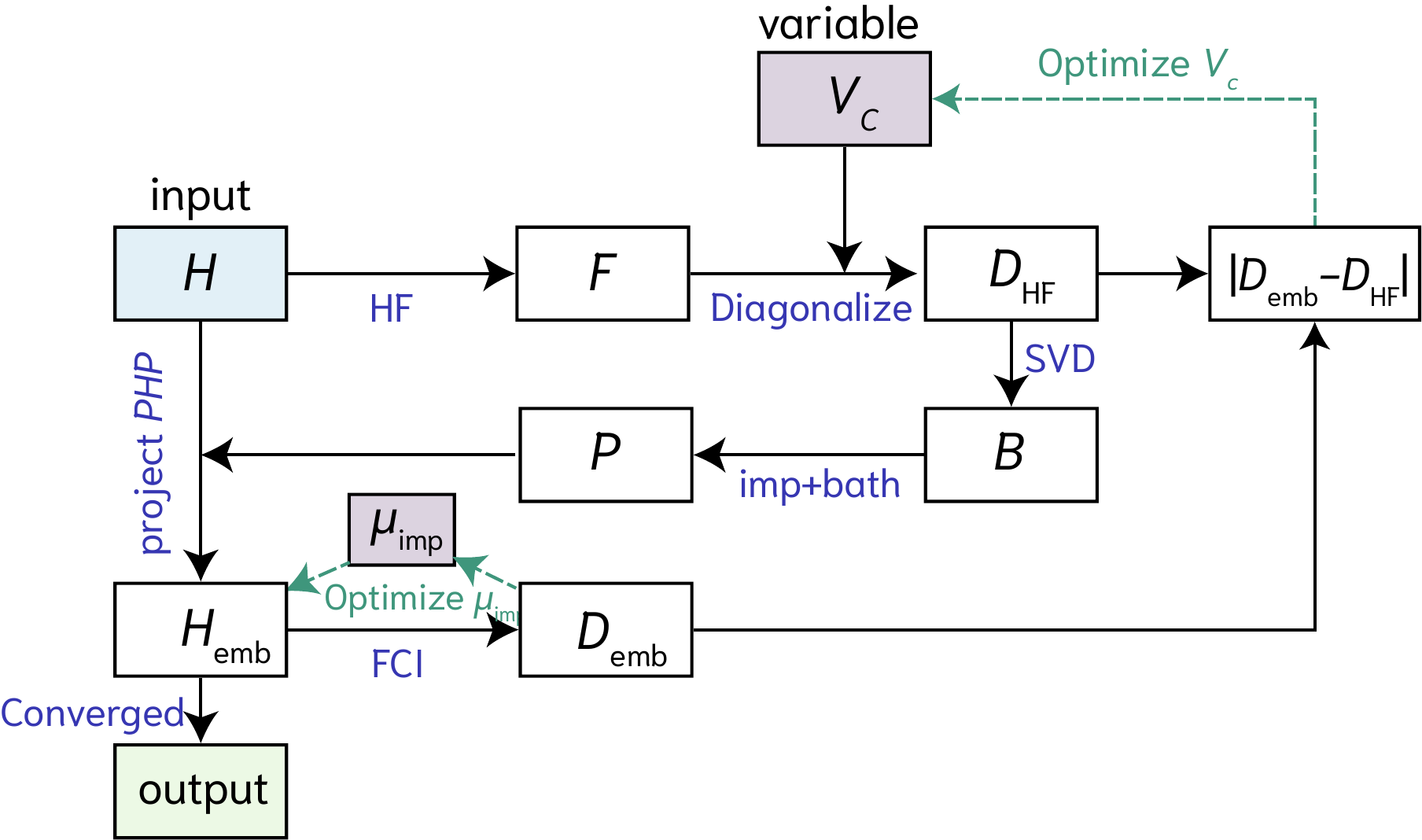}
    \caption{DMET algorithm. Dashed green arrows denote optimization loops. $H$ is the input lattice Hamiltonian, and $V_c$ is the correlation potential to be optimized. $F$ denotes the Fock operator, and $D_\text{HF}$ is the mean-field 1RDM. $B$ labels the bath orbitals, and $P$ is the projector onto the embedding space. $H_\text{emb}$ is the embedding Hamiltonian constructed with an interacting bath, and $D_\text{emb}$ is the corresponding embedding 1RDM. $\mu_\text{imp}$ is the impurity chemical potential optimized in the inner loop. The final output is the converged DMET solution after all loops have converged.}
    \label{fig:dmet_algorithm}
\end{figure}

We present the DMET algorithm in FIG~\ref{fig:dmet_algorithm}.

\begin{figure}[t!]
\centering

\begin{subcaptionbox}{1D Hydrogen Chain at $\beta = 100 E_\text{h}^{-1}$\label{sub:si_conv1d_T001}}{\includegraphics[width=\linewidth]{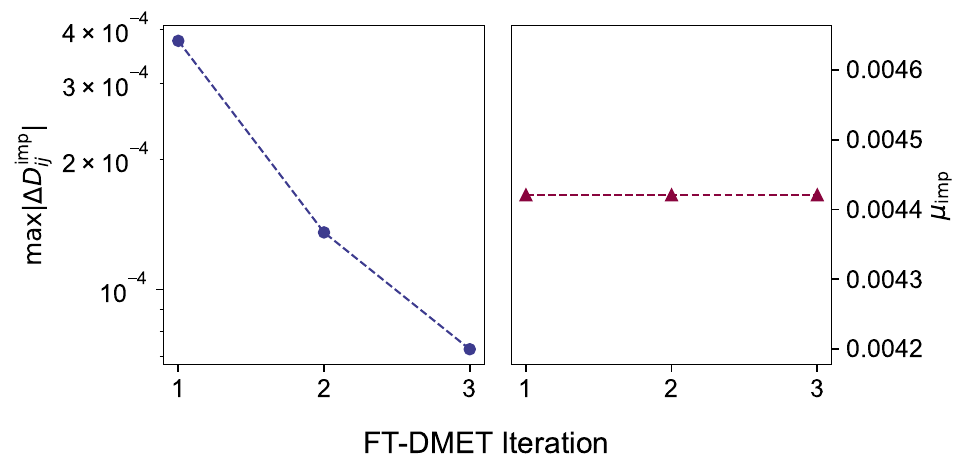}}
\end{subcaptionbox}

\vspace{0.5em}

\begin{subcaptionbox}{1D Hydrogen Chain at $\beta = 50 E_\text{h}^{-1}$\label{sub:si_conv1d_T002}}{\includegraphics[width=\linewidth]{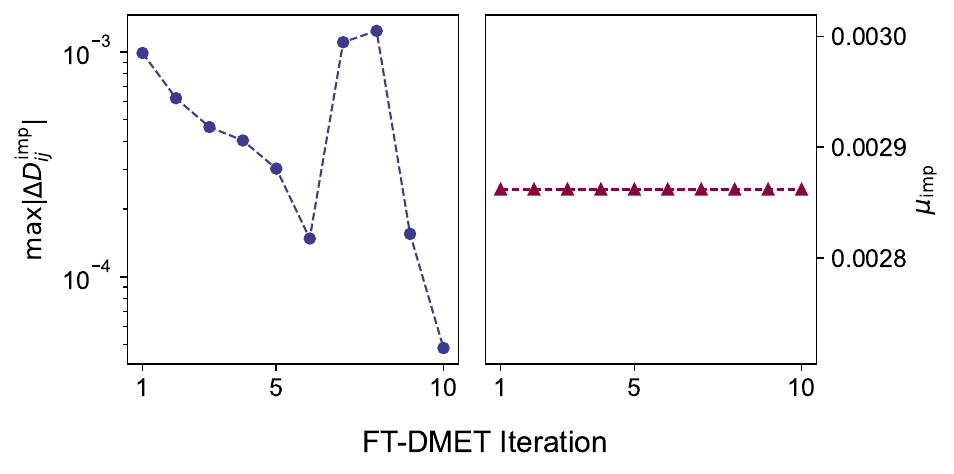}}
\end{subcaptionbox}

\vspace{0.5em}

\begin{subcaptionbox}{2D Hydrogen Lattice at $\beta = 100 E_\text{h}^{-1}$\label{sub:si_conv2d_T001}}{\includegraphics[width=\linewidth]{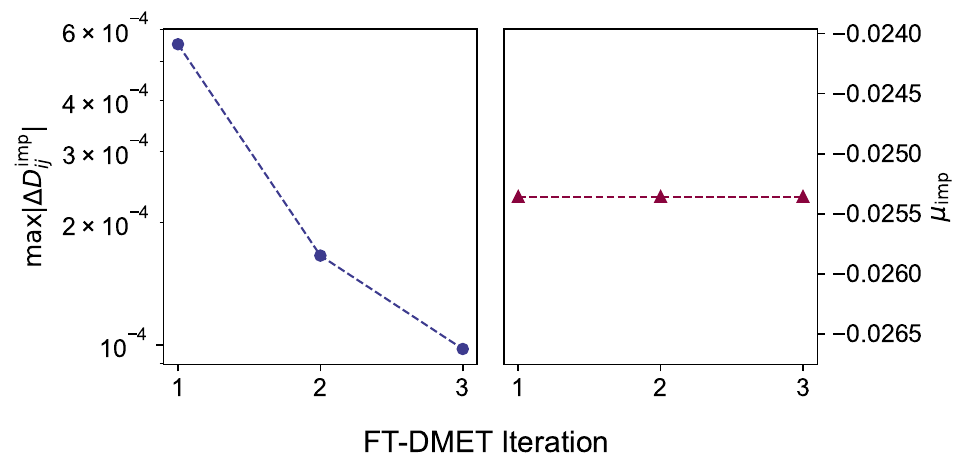}}
\end{subcaptionbox}

\vspace{0.5em}

\begin{subcaptionbox}{2D Hydrogen Lattice at $\beta = 50 E_\text{h}^{-1}$\label{sub:si_conv2d_T002}}{\includegraphics[width=\linewidth]{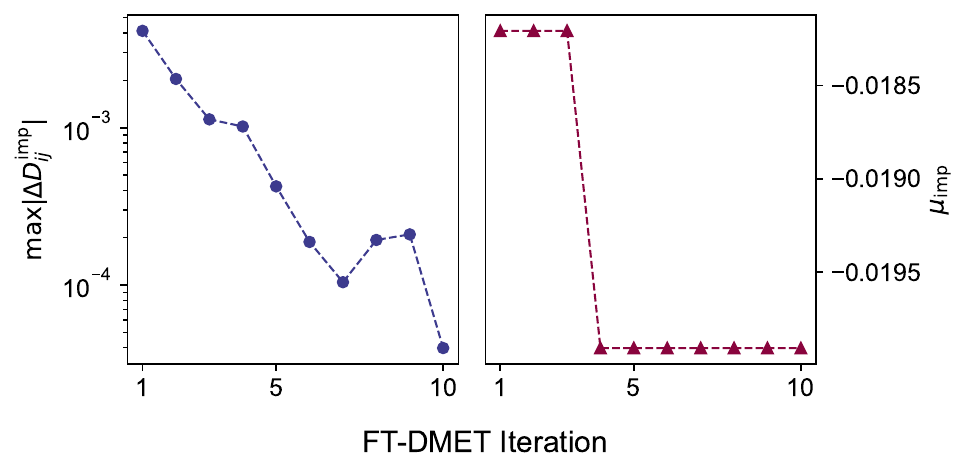}}
\end{subcaptionbox}

\caption{{FT-DMET convergence for (a)-(b) one-dimensional hydrogen chain and (c)-(d) two-dimensional hydrogen square lattice systems.}
}
\label{fig:si_dmet_convergence}
\end{figure}

\section{FT-DMET Convergence}

We provide convergence details for the full SCF procedure in FT-DMET. The converged ground-state solution is used as the initial guess for the finite-temperature calculations. This includes the mean-field 1RDM used to initialize the $k$-space mean-field calculation, the impurity chemical potential $\mu_\text{imp}$ for the impurity solver, and the correlation potential for the FT-DMET SCF loop.

We examine a one-dimensional hydrogen chain at $R = 1.5\,a_0$ with a four-atom impurity and the STO-6G basis set, as well as a two-dimensional hydrogen square lattice at $R = 1.5\,a_0$ with a $2 \times 2$ impurity cluster in the same basis. We consider temperatures corresponding to $\beta = 100$ and $50\,E_\text{h}^{-1}$. The convergence criteria are set to $|\Delta E^{\text{imp}}| \le 10^{-6}\,E_{\text{h}}$ and $\max|\Delta D^{\text{imp}}_{ij}| \le 10^{-4}$, where $\Delta$ denotes the difference between two consecutive iterations.

Both systems exhibit similar convergence behavior. For $\beta = 100\,E_\text{h}^{-1}$, the calculations converge within three iterations, while for $\beta = 50\,E_\text{h}^{-1}$, convergence is reached after about ten iterations. In Fig.~\ref{fig:si_dmet_convergence}, we show the convergence profiles of $\max|\Delta D^{\text{imp}}_{ij}|$ and $\mu_\text{imp}$ as a function of the iteration number. The results indicate rapid convergence at the lower temperature ($\beta = 100\,E_\text{h}^{-1}$) and slightly slower convergence at the higher temperature ($\beta = 50\,E_\text{h}^{-1}$). Meanwhile, the impurity chemical potential $\mu_{\text{imp}}$ remains nearly unchanged throughout the iterations, suggesting that the inner DMET loop can be truncated after the first iteration.

To further reduce the number of DMET iterations at higher temperatures, one can initialize the calculation from the solution obtained at a lower temperature. In practice, finite-temperature simulations often evaluate a grid of temperatures to track the change of order parameters. In such cases, this one-shot FT-DMET strategy provides an efficient way to obtain accurate results at higher temperatures.

\bibliography{references}

\end{document}